\documentclass{emulateapj}
\usepackage{graphicx}
\usepackage[space]{grffile}
\usepackage{latexsym}
\usepackage{textcomp}
\usepackage{longtable}
\usepackage{multirow,booktabs}
\usepackage{amsfonts,amsmath,amssymb}
\usepackage{natbib}
\usepackage{url}
\usepackage{CJKutf8}
\usepackage{hyperref}
\usepackage{enumerate}
\hypersetup{colorlinks=false,pdfborder={0 0 0}}
\newif\iflatexml\latexmlfalse
\usepackage[utf8]{inputenc}
\usepackage[english]{babel}
\usepackage{graphicx, color}

\newcommand\T{\rule{0pt}{2.6ex}}       

\newcommand{\be}{\begin{equation}}
\newcommand{\ee}{\end{equation}}
\newcommand*{\teff}{$T_{\rm eff}$}
\newcommand*{\logg}{$\log~g$}
\newcommand*{\feh}{[Fe/H]}

\newcommand*{\aM}{[$\alpha$/M]}

\definecolor{todo}{RGB}{200,0,0}

\shorttitle{SDSS Data Release 16}
\shortauthors{SDSS-IV Collaboration}

\begin{document}

\title{The Sixteenth Data Release of the Sloan Digital Sky Surveys:  First Release from the APOGEE-2 Southern Survey and Full Release of eBOSS Spectra}

\email{spokesperson@sdss.org}
\author{Romina Ahumada\altaffilmark{1},
Carlos Allende Prieto\altaffilmark{2,3},
Andr\'es Almeida\altaffilmark{4},
Friedrich Anders\altaffilmark{5,6},
Scott F. Anderson\altaffilmark{7},
Brett H. Andrews\altaffilmark{8},
Borja Anguiano\altaffilmark{9},
Riccardo Arcodia\altaffilmark{10},
Eric Armengaud\altaffilmark{11},
Marie Aubert\altaffilmark{12}
% ...
}
\altaffiltext{1}{Instituto de Astronom\'ia, Universidad Cat\'olica del Norte, Av. Angamos 0610, Antofagasta, Chile}
\altaffiltext{2}{Instituto de Astrof\'{\i}sica de Canarias (IAC), C/ Via L\'actea s/n, E-38205 La Laguna, Tenerife, Spain}
\altaffiltext{3}{Universidad de La Laguna (ULL), Departamento de Astrof\'{\i}sica, E-38206 La Laguna, Tenerife Spain}
\altaffiltext{4}{Instituto de Investigaci\'on Multidisciplinario en Ciencia y Tecnolog\'ia, Universidad de La Serena. Avenida Ra\'ul Bitr\'an S/N, La Serena, Chile}
\altaffiltext{5}{Leibniz-Institut f\"ur Astrophysik Potsdam (AIP), An der Sternwarte 16, D-14482 Potsdam, Germany}
\altaffiltext{6}{Institut de Ci\`{e}ncies del Cosmos, Universitat de Barcelona (IEEC-UB), Carrer Mart\'{i} i Franqu\`{e}s 1, 08028 Barcelona, Spain}
\altaffiltext{7}{Department of Astronomy, University of Washington, Box 351580, Seattle, WA 98195, USA}
\altaffiltext{8}{PITT PACC, Department of Physics and Astronomy, University of Pittsburgh, Pittsburgh, PA 15260, USA}
\altaffiltext{9}{Department of Astronomy, University of Virginia, Charlottesville, VA 22904-4325, USA}
\altaffiltext{10}{Max-Planck-Institut f\"ur extraterrestrische Physik, Gie{\ss}enbachstra{\ss}e 1, 85748 Garching, Germany}
\altaffiltext{11}{IRFU, CEA, Universit\'e Paris-Saclay, F91191 Gif-sur-Yvette, France}
\altaffiltext{12}{Aix Marseille Universit\'e, CNRS/IN2P3, CPPM, Marseille, France}
% ...

\begin{abstract}
This paper documents the sixteenth data release (DR16) from the Sloan Digital Sky Surveys; the fourth and penultimate from the fourth phase (SDSS-IV). This is the first release of data from the southern hemisphere survey of the Apache Point Observatory Galactic Evolution Experiment 2 (APOGEE-2); new data from APOGEE-2 North are also included. DR16 is also notable as the final data release for the main cosmological program of the Extended Baryon Oscillation Spectroscopic Survey (eBOSS), and all raw and reduced spectra from that project are released here. DR16 also includes all the data from the Time Domain Spectroscopic Survey (TDSS) and new data from the SPectroscopic IDentification of ERosita Survey (SPIDERS) programs, both of which were co-observed on eBOSS plates. DR16 has no new data from the Mapping Nearby Galaxies at Apache Point Observatory (MaNGA) survey (or the MaNGA Stellar Library ``MaStar"). We also preview future SDSS-V operations (due to start in 2020), and summarize plans for the final SDSS-IV data release (DR17).  
\end{abstract}


\section{Introduction}
The Sloan Digital Sky Surveys (SDSS) have been observing the skies
from Apache Point Observatory (APO) since 1998 (using the 2.5m Sloan Foundation Telescope, \citealt{2006AJ....131.2332G}) and from Las Campanas Observatory (LCO) since 2017 (using the du Pont 2.5m Telescope).

Representing the fourth phase of SDSS, SDSS-IV \citep{2017AJ....154...28B} consists
of three main surveys; the Extended Baryon Oscillation Spectroscopic
Survey (eBOSS; \citealt{Dawson16}), Mapping Nearby Galaxies
at APO (MaNGA; \citealt{2015ApJ...798....7B}), and the APO Galactic
Evolution Experiment 2 (APOGEE-2; \citealt{Majewski2017}). Within eBOSS,
SDSS-IV has also conducted two smaller programs: the SPectroscopic
IDentification of ERosita Sources (SPIDERS; \citealt{Clerc2016,
Dwelly17}) and the Time Domain Spectroscopic Survey
(TDSS; \citealt{morganson15a}). These programs have investigated a
broad range of cosmological scales, including cosmology with large-scale structure in
eBOSS, the population of quasars and variable or X-ray-emitting stars
with TDSS and SPIDERS; nearby galaxies in MaNGA; and the Milky Way and
its stars in APOGEE-2.

This paper documents the sixteenth data release from SDSS (DR16), the latest in a
series that began in 2001 (\citealt{2002AJ....123..485S}). It is the fourth data release from SDSS-IV (following DR13:  \citealt{2017ApJS..233...25A}; DR14: \citealt{2018ApJS..235...42A} and DR15: \citealt{2019ApJS..240...23A}). A complete overview of the scope of DR16 is provided in \S \ref{sec:scope}, and information on how to access the data can be found in \S \ref{sec:access}.  DR16 contains three
important milestones:
\begin{enumerate}
\item The first data from APOGEE-2 South (APOGEE-2S), which is mapping the
Milky Way in the Southern hemisphere from the du Pont Telescope at LCO. With SDSS now operating APOGEE instruments in two hemispheres,
all major components of the Milky Way are accessible (see \S \ref{sec:apogee})
\item The first and final release of eBOSS spectra from the emission line galaxy (ELG)
cosmology program.  The entirety of this large-scale structure survey was conducted in the interval
between DR14 and DR16. Covering the redshift range $0.6<z<1.1$, the eBOSS ELG program
represents the highest redshift galaxy survey ever conducted within SDSS.
\item The full and final release of spectra from the main observing program of eBOSS, completing that cosmological redshift
survey. DR16 therefore marks the end of a twenty-year stretch during
which SDSS performed a redshift survey of the large-scale structure in the
universe.  Over this span, SDSS produced a catalog of spectroscopic galaxy
redshifts that is a factor of more than five larger than any other program. 
DR16 provides spectra along with usable redshifts for around 2.6 million
unique galaxies. The catalogues that contain the information to accurately measure the clustering statistics of ELGs, luminous red galaxies (LRGs), quasars, and Lyman-$\alpha$ absorption will be released later (see \S \ref{sec:eboss}).
\end{enumerate}

DR16 also represents the full release of the TDSS subprogram, which in total releases spectra for 131,552 variable sources (see \S \ref{sec:tdss}). The SPIDERS subprogram will have a small number of observations in the future to cover eROSITA targets, but DR16 releases a number of Value Added Catalogs (VACs) characterizing both X-ray cluster and X-ray point sources that have already been observed (as well as the optical spectra; see \S \ref{sec:spiders}). There are no new data from MaNGA or MaStar \citep{Yan2019} in DR16; however, a number of new or updated VACs based on DR15 MaNGA data are released (see \S \ref{sec:manga}).

\section{Scope of DR16}
\label{sec:scope}
Following the tradition of previous SDSS data releases, DR16 is a cumulative data release. This means that all previous data releases are included in DR16, and data products and catalogs of these previous releases will remain accessible on our data servers. Table \ref{tab:scope} shows the number of spectra contained in DR16 along with those from previous releases and demonstrates the incremental gains with each release. We strongly advise to always use the most recent SDSS data release, as data will have been reprocessed using updated data reduction pipelines, and catalogs may have been updated with new entries and/or improved analysis methods. These changes between DR16 and previous data releases are documented in this paper and on the DR16 website \url{https://www.sdss.org/dr16}.

\begin{deluxetable*}{llrrrr}
\tablewidth{6.5in}
\tablecaption{SDSS-IV spectroscopic data in DR13--DR16 \label{tab:scope}} 
\tablehead{ 
\colhead{Survey} & \colhead{Target Category} & \colhead{DR13} & \colhead{DR14}  & \colhead{DR15} & \colhead{DR16}}
\startdata
{eBOSS} \\
& \multicolumn{1}{r}{LRG samples} & 32968 & 138777 & 138777 & 298762 \\
&  \multicolumn{1}{r}{ELG samples} & 14459 & 35094 & 35094 & 269889 \\
& \multicolumn{1}{r}{Main QSO Sample}  & 33928 & 188277 & 188277 & 434820 \\	
& \multicolumn{1}{r}{Variability Selected QSOs} & 22756 & 87270 & 87270 & 185816 \\
& \multicolumn{1}{r}{Other QSO samples} & 24840 & 43502  & 43502 & 70785 \\
& \multicolumn{1}{r}{TDSS Targets} & 17927 & 57675 & 57675  & 131552\\
& \multicolumn{1}{r}{SPIDERS Targets} &  3133 & 16394 & 16394  & 36300\\
& \multicolumn{1}{r}{Reverberation Mapping} &   849\tablenotemark{a}  & 849\tablenotemark{a}  & 849\tablenotemark{a} & 849\tablenotemark{a}  \\
& \multicolumn{1}{r}{Standard Stars/White Dwarfs} &  53584 & 63880 & 63880 & 84605  \\
\tableline 
{APOGEE-2} \T \\
&\multicolumn{1}{r}{Main Red Star Sample}  & 109376 & 184148 & 184148 & 281575\\
&\multicolumn{1}{r}{AllStar Entries}  & 164562 & 277371 & 277371 & 473307\tablenotemark{b} \\
&\multicolumn{1}{r}{APOGEE-2S Main Red Star Sample}  & - & - & - &56480  \\
&\multicolumn{1}{r}{APOGEE-2S AllStar Entries}  & - & - & - & 102200 \\
&\multicolumn{1}{r}{APOGEE-2S Contributed AllStar Entries}  & - & - & - & 37409 \\
&\multicolumn{1}{r}{NMSU 1-meter AllStar Entries}  & 894 & 1018 & 1018 & 1071 \\
&\multicolumn{1}{r}{Telluric AllStar Entries} & 17293 &  27127 &  27127 & 34016 \\
&\multicolumn{1}{r}{APOGEE-N Commissioning stars}  & 11917  & 12194 & 12194 & 12194  \\
\tableline 
MaNGA \\
&\multicolumn{1}{l}{MaNGA Cubes} &  1390 &  2812 &  4824 & 4824 \\ 
& \multicolumn{4}{l}{MaNGA main galaxy sample: } \\ 
& \multicolumn{1}{r}{\tt PRIMARY\_v1\_2} &  600  & 1278  & 2126 & 2126 \\ 
&  \multicolumn{1}{r}{\tt SECONDARY\_v1\_2} &  473  & 947  & 1665 & 1665 \\ 
& \multicolumn{1}{r}{\tt COLOR-ENHANCED\_v1\_2} & 216  & 447 & 710 &  710 \\  
& \multicolumn{1}{l}{MaStar (MaNGA Stellar Library)} & - & - & 3326 & 3326 \\
& \multicolumn{1}{l}{Other MaNGA ancillary targets\tablenotemark{c}} &  31 & 121 & 324 & 324 \\
\vspace{-0.1in}
\tablenotetext{a}{The number of RM targets remains the same, but the number of visits increases.} 
\tablenotetext{b}{This number includes multiple entries for some stars; there are 437,485 unique stars.}
\tablenotetext{c}{Many MaNGA ancillary targets were also observed as part of the main galaxy sample, and are counted twice in this table; some ancillary targets are not galaxies.} 
\enddata
\end{deluxetable*}

The content of DR16 is given by the following sets of data products:

\begin{enumerate}

\item eBOSS is releasing 860,935 new optical spectra of galaxies and quasars with respect to its previous SDSS data release. These targets were observed between MJD 57520 (May 11th 2016) and 58543 (March 1st 2019), and bring the total number of spectra observed by eBOSS to 1.4 million. This number includes spectra observed as part of the TDSS and SPIDERS sub-surveys, as well as the spectra taken as part of the eBOSS reverberation mapping ancillary program. All spectra, whether released previously or for the first time in this data release, have been processed using the latest version of the eBOSS data reduction pipeline {\tt v5\_13\_0}. In addition to the spectra, eBOSS is also releasing catalogs of redshifts, as well as various value-added catalogs (VACs; see Table \ref{table:vac}). DR16 is the last SDSS data release that will contain new eBOSS spectra from the main program, as this survey has now finished.  Additional observations of X-ray sources under the SPIDERS program and continued monitoring of quasars under the reverberation mapping program are planned before the end of SDSS-IV, which will lead to another increment of single-fiber spectra from the Baryon Oscillation Spectroscopic Survey (BOSS) spectrograph in DR17.

\item APOGEE-2 is including 751,864 new infrared spectra;\footnote{the number of entries in the All Visit file, which is larger than the number of combined spectra having entries in the AllStar file as listed in Table \ref{tab:scope}} the new spectra comprise both observations of 195,936 new stars and additional epochs on targets included in previous DRs. The majority of the stars are in the Milky Way galaxy (including Omega Centauri), but DR16 also contains stars from, the Large and Small Magellanic Clouds, and dwarf Spheroidal satellites. A total of 262,997 spectra, for 102,200 unique stars, were obtained in the southern hemisphere from the {APOGEE-S} spectrograph at LCO. These new spectra were obtained from MJD 57643 to MJD 58301 (September 12th 2016 to July 2nd 2018) for APOGEE-2N from APO and from MJD 57829 to MJD 58358 (March 17, 2017 to August 28, 2018) for APOGEE-2S from LCO. DR16 also includes all previously released APOGEE and APOGEE-2 spectra, which have been re-reduced with the latest version of the APOGEE data reduction and analysis pipeline. In addition to the reduced spectra, element abundances and stellar parameters are included in this data release. APOGEE-2 is also releasing a number of VACs (Table \ref{table:vac})

\item MaNGA and MaStar are not releasing any new spectra in this data release; the spectra and data products included in DR16 are therefore identical to the ones that were released in DR15. However, MaNGA is contributing a number of of new or updated VACs in DR16, which are based on the DR15 sample and data products (see Table \ref{table:vac}).

\item Since SDSS data releases are cumulative, {\bf DR16 also includes data from all previous SDSS data releases.} All BOSS and eBOSS, APOGEE and APOGEE-2 spectra that were previously released have all been reprocessed with the latest reduction and analysis pipelines. The MaNGA and MaStar data in DR16 are identical to those in DR15 \citep{2019ApJS..240...23A}; SDSS-III MARVELS spectra have not changed since DR12 \citep{2015ApJS..219...12A}. SDSS Legacy Spectra in DR16 are the same as those released in their final form in DR8 \citep{2011ApJS..193...29A}, and the SEGUE-1 and SEGUE-2 survey data in DR16 are identical to the final reductions released with DR9 \citep{2012ApJS..203...21A}. The SDSS imaging had its most recent change in DR13 \citep{2017ApJS..233...25A}, when it was recalibrated for eBOSS imaging purposes and DR16 contains this version of the imaging. 

\end{enumerate}

An overview of the total spectroscopic content of DR16, with number of spectra included, is given in Table \ref{tab:scope}. An overview of the value-added catalogs that are new or updated in DR16 can be found in Table \ref{table:vac}; adding these to the VACs previously released in SDSS, there are a total of 46 VACs in DR16\footnote{That is 40 previous released VACs, 7 of which are updated in DR16, and 6 VACs new to DR16.}.

\begin{deluxetable*}{lll}
\tablecaption{New or Updated Value Added Catalogs (VACs) \label{table:vac}}
\tablehead{\colhead{Description} & \colhead{Section} &  \colhead{Reference(s)}}
\startdata
APOGEE-2 Red Clumps  & \S \ref{vac:rc}  & \citet{2014ApJ...790..127B}\\
APOGEE-2 \texttt{astroNN}  & \S \ref{vac:astroNN} & \citet{2019MNRAS.483.3255L}\\
APOGEE-2 \textit{Joker} & \S \ref{vac:joker} &  \citet{PriceWhelan2017, PriceWhelan2018, PriceWhelan2020} \\
APOGEE-2 OCCAM & \S \ref{vac:occam} &  \citet{Donor2018,Donor2020} \\
APOGEE-2 StarHorse & \S \ref{vac:starhorse} & \citet{2018MNRAS.476.2556Q, Anders2019}; \\
& & \citet{Quieroz2019} \\
eBOSS ELG classification  & \S \ref{vac:eboss}& \citet{Zhang2019} \\
SDSS Galaxy Single Fiber FIREFLY & \S \ref{vac:eboss} & \citet{Comparat2017} \\
SPIDERS X-ray clusters & \S \ref{vac:clusters} & \citet{Clerc2016}, C. Kirkpatrick et al. in prep.\\
SPIDERS Rosat and XMM\tablenotemark{a}-Slew Sources & \S \ref{vac:agn} & \citet{Comparat2020}  \\
SPIDERS Multiwavelength Properties of RASS and XMMSL AGN & \S \ref{vac:rass} & \citet{Comparat2020}\\
SPIDERS Black Hole Masses  & \S \ref{vac:coffey} & \citet{Coffey2019}\\
MaNGA Stellar Masses from PCA  & \S \ref{vac:pca} & \citet{Pace2019a,Pace2019b} \\
MaNGA {\tt PawlikMorph} & \S \ref{vac:pawlikmorph} & \citet{Pawlik2016}
\enddata
\tablenotetext{a}{X-ray Multi-Mirror Mission}
\end{deluxetable*}

\section{Data Access}
\label{sec:access}
The SDSS data products included in DR16 are publicly available through several different channels. The best way to access the data products depends on the particular product, and the goal of the user. The different access options are described on the SDSS website \url{https://www.sdss.org/dr16/data_access/}, and we also describe them below. We provide a variety of tutorials and examples for accessing data products online at \url{https://www.sdss.org/dr16/tutorials/}.

All software that was used by SDSS to reduce and process data, as well as to construct derived data products is publicly available in either SVN or Github repositories; an overview of available software and where to retrieve it is given on \url{https://www.sdss.org/dr16/software/}.

\subsection{Science Archive Server; SAS}

The main path to access the raw and reduced imaging and spectroscopic data directly, as well as obtain intermediate data products and value-added catalogs, is through the SDSS Science Archive Server (SAS, \url{https://data.sdss.org/sas/}). Note that all previous data releases are also available on this server, but we recommend that users always adopt the latest data release, as these are reduced with the latest versions of the data reduction software. The SAS is a file-based system, which allows data downloads by browsing or through tools such as {\tt rsync}, {\tt wget} and Globus Online (see \url{https://www.sdss.org/dr16/data_access/bulk} for more details). The content of each data product on the SAS is outlined in its data model, which can be accessed through \url{https://data.sdss.org/datamodel/}.

\subsection{Science Archive Webapp; SAW}

Most of the reduced images and spectra on the SAS are also accessible through the Science Archive Webapp (SAW), which provides the user with options to display spectra and overlay model fits. The SAW includes search options to access specific subsamples of spectra, e.g. based on coordinates, redshift and/or observing programs. Searches can also be saved as ``permalinks" to allow sharing with collaborators and future use. Links are provided to download the spectra directly from the SAS, and to open SkyServer Explore pages for the objects displayed (see below for a description of the SkyServer). The SAW contains imaging, optical single-fiber spectra (SDSS-I/II, SEGUE, BOSS, eBOSS), infrared spectra (APOGEE-1/2) and stellar spectra of the MaStar stellar library. All of these webapps are linked from \url{https://dr16.sdss.org/}. Just like the SAS, the SAW provides access to previous data releases (back to DR8).

\subsection{Marvin for MaNGA}

Integral-field spectroscopic data (MaNGA) are not available in the SAW because they follow a different data format from the single object spectra. Instead, the MaNGA data can be accessed through Marvin (\url{https://dr16.sdss.org/marvin/}; \citealt{2019AJ....158...74C}). Marvin can be used to both visualize and analyze MaNGA data products and perform queries on MaNGA meta-data remotely. Marvin also contains a suite of Python tools, available through pip-install, that simplify interacting with the MaNGA data products and meta-data. More information, including installation instructions for Marvin, can be found here: \url{https://sdss-marvin.readthedocs.io/en/stable/}. In DR16, although no new MaNGA data products are included, Marvin has been upgraded by providing access to a number of MaNGA value-added catalogs based on DR15 data.

\subsection{Catalog Archive Server, CAS}

The SDSS catalogs can be found and queried on the Catalog Archive Server or CAS \citep{2008CSE....10...30T}. These catalogs contain photometric and spectroscopic properties, as well as derived data products. Several value added catalogs are also available on the CAS. For quick inspection of objects or small queries, the SkyServer webapp (\url{https://skyserver.sdss.org}) is the recommended route to access the catalogs: it contains amongst other facilities the Quick Look and Explore tools, as well as the option for SQL queries in synchronous mode directly in the browser. The SkyServer also contains tutorials and examples of SQL syntax (\url{http://skyserver.sdss.org/public/en/help/docs/docshome.aspx}). For larger queries, CASJobs (\url{https://skyserver.sdss.org/casjobs}) should be used, as it allows for asynchronous queries in batch mode. Users of CASJobs will need to create a (free of cost) personal account, which comes with storage space for query results \citep{2008CSE....10...18L}. A third way to access the SDSS catalogs is through the SciServer (\url{https://www.sciserver.org}), which is integrated with the CAS. SciServer allows users to run Jupyter notebooks in Docker containers, amongst other services.  

\subsection{Data Access for Education}

We are providing access to a growing set of Jupyter Notebooks that have been developed for introductory\footnote{\url{https://github.com/ritatojeiro/SDSSEPO}} and upper-level\footnote{\url{https://github.com/brittlundgren/SDSS-EPO}} university astronomy laboratory courses. These Python-based activities are designed to be run on the SciServer platform\footnote{\url{http://www.sciserver.org/}}, which enables the analysis and visualization of the vast SDSS dataset from a web browser, without requiring any additional software or data downloads.

Additionally, Voyages (\url{http://voyages.sdss.org/}) provides activities and resources to help younger audiences explore the SDSS data. Voyages has been specifically developed to be used in secondary schools, and contains pointers to K-12 US science standards.  A Spanish language version of these resources is now available at \url{http://voyages.sdss.org/es}

\section{APOGEE-2: First Release of Southern Hemisphere Data, and More from the North} 

\label{sec:apogee}

APOGEE is performing a chemodynamical investigation across the entire Milky Way Galaxy with two similarly designed near-infrared, high-resolution multiplexed spectrographs. DR16 constitutes the fifth release of data from APOGEE, which has run in two phases (APOGEE-1 and APOGEE-2) spanning both SDSS-III and SDSS-IV.  For approximately 3 years (August 2011- July 2014), APOGEE-1 survey observations were conducted at the 2.5m Sloan Foundation Telescope at APO as part of SDSS-III. In August 2014, at the start of SDSS-IV, APOGEE-2 continued data acquisition at the APO northern hemisphere site (APOGEE-2N).  With the build of a second spectrograph (\citealt{Wilson2019}), APOGEE-2 commenced southern hemisphere operations at the 2.5m Ir\'en\'e du Pont Telescope at LCO (APOGEE-2S) in April 2017.  \citet{Majewski2017} provides an overview of the APOGEE-1 Survey (with a forthcoming S. Majewski et al. in prep. planned to provide an an overview of the APOGEE-2 Survey).

In detail, the APOGEE data in DR16 encompasses all SDSS-III APOGEE-1 data and SDSS-IV APOGEE-2 data acquired with both instruments through August 2018.  The current release includes two additional years of APOGEE-2N data and almost doubles the number of stars with available spectra as compared to the previous public release \citep[in DR14: ][]{2018ApJS..235...42A}.  DR16 presents the first 16 months of data from APOGEE-2S. Thus, DR16 is the first release from APOGEE that includes data from across the entire night sky. 

\begin{figure*}
\centering
\includegraphics[angle=0,width=15cm]{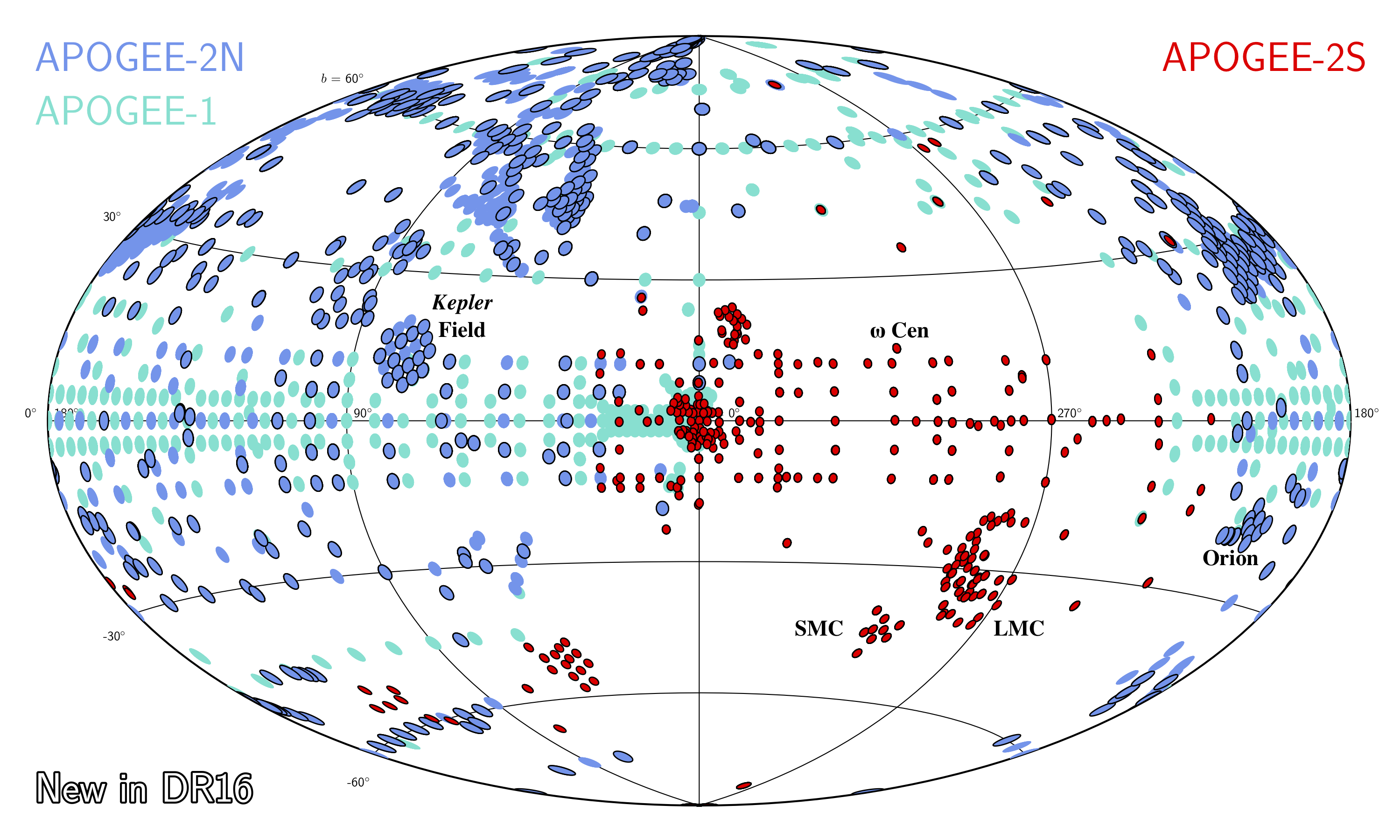}
\caption{DR16 APOGEE sky coverage in Galactic coordinates. Each symbol represents a field, which is 7 square degrees for APOGEE-1 in cyan and APOGEE-2N in blue and 2.8 square degrees for APOGEE-2S in red (this difference is due to the different field-of-view of the two telescopes; see \S \ref{apogeeS}). Fields that have new data presented in DR16 are hi-lighted with a black outline. }
\label{fig:apogeedr16}
\end{figure*}

\begin{figure*}
\centering
\includegraphics[angle=0,width=15cm]{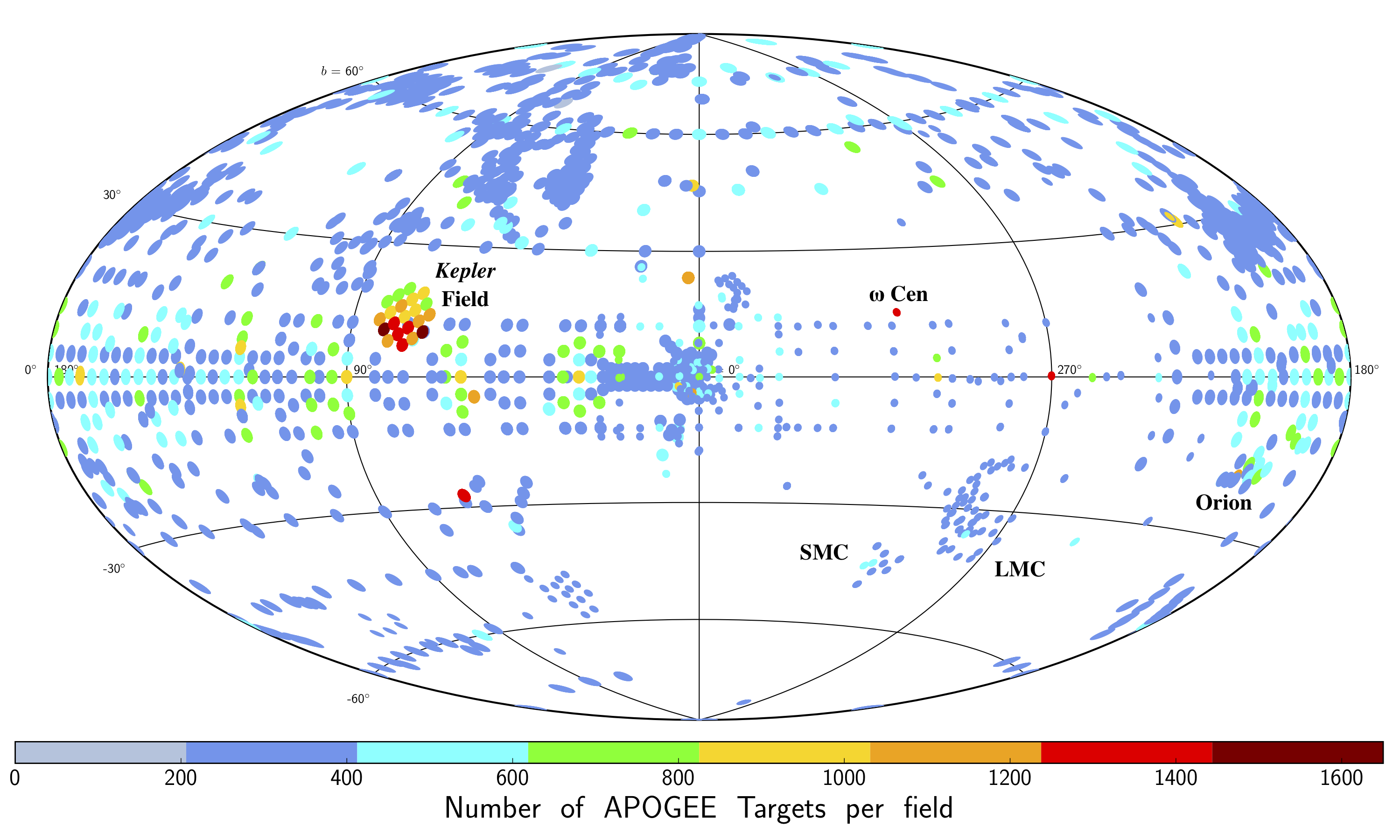}
\caption{A sky map in Galactic coordinates showing the number of stars per APOGEE field (across APOGEE-1, 2N, and 2S). The disk is targeted with a symmetric dense grid within $|b| < 15\deg$. The dense coverage of the bulge and inner Galaxy is for $l < 30 \deg$. Other special programs, like the {\it Kepler-2} follow-up, have initial data in DR16. The circle sizes reflect the different field-of-view of APOGEE-N and APOGEE-S; see \S \ref{apogeeS}.}
\label{fig:apogeenstars}
\end{figure*}

DR16 contains APOGEE data and information for 437,485 unique stars, including reduced and visit-combined spectra, radial velocity information, atmospheric parameters, and individual element abundances; nearly 1.8 million individual visit spectra are included. Figure \ref{fig:apogeedr16} displays the APOGEE DR16 coverage in Galactic coordinates where each point represents a single ``field" and is color-coded by the overall survey component (e.g., APOGEE, APOGEE-2N, and APOGEE-2S). Fields newly released in DR16 are encircled with black. As shown in this figure, the dual hemisphere view of APOGEE allows for targeting of all Milky Way components: the inner and outer halo, the four disk quadrants, and the full expanse of the bulge.  The first APOGEE-2S observations of various Southern Hemisphere objects, such as Omega Centauri ($l,b = 309^\circ, 15^\circ$) and our current targetting of the Large and Small Magellanic Clouds ($l,b = 280^\circ, -33^\circ$ and $303^\circ, -44^\circ$ respectively), are visible in Figure \ref{fig:apogeedr16} as high density areas of observation.  Moreover, DR16 features substantially increased coverage at high Galactic latitudes as APOGEE continues to piggy-back on MaNGA-led observing during dark time. Figure \ref{fig:apogeenstars} has the same projection, but uses color-coding to convey the number of unique targets for each of the APOGEE fields. Particularly dense regions include the Kepler field which serves multiple scientific programs, as well as APOGEE ``deep" fields observed with multiple ``cohorts" (see \citealt{Zasowski_2017_apogee2targeting}). Detailed discussions of our targeting strategies for each Galactic component, as well as an evaluation of their efficacy, will be presented in forthcoming focused papers (R. Beaton et al. in prep, F. Santana et al. in prep).

\subsection{APOGEE Southern Survey Overview} \label{apogeeS}

The APOGEE-2S Survey has been enabled by the construction of a second APOGEE spectrograph.  The second instrument is a near duplicate of the first with comparable performance, simultaneously delivering 300 spectra in the $H$-band wavelength regime ($\lambda$ = 1.5$\mu$m to 1.7$\mu$m) at a resolution of $R\sim 22,500$.  Slight differences occur between the two instruments with respect to image quality and resolution across the detectors as described in detail in \citet{Wilson2019}.

The telescopes of the Northern and Southern Hemisphere sites have the same apertures.  However, because the du Pont telescope was designed with a slower focal ratio (f/7.5) than the Sloan Foundation telescope (f/5), the resulting field-of-view for APOGEE-2S is smaller than APOGEE-2N and the fibers subtend a smaller angular area. The difference in field-of-view is evident in Figure \ref{fig:apogeedr16} by comparing the size of the red points (LCO fields) to those shown in blue or cyan (APO fields). However, the image quality (seeing) at LCO is generally better than that at APO, and this roughly compensates for the smaller angular diameter fibers such that the typical throughput at LCO is similar to, or even better than, that obtained at APO.

\subsection{General APOGEE Targeting}
Extensive descriptions of the target selection and strategy are found in \citet{Zasowski2013} for APOGEE-1 and in \citet{Zasowski_2017_apogee2targeting} for APOGEE-2. Details about the final selection method used for APOGEE-2N and APOGEE-2S will be presented in R. Beaton, et. al in prep. and F. Santana et al. in prep, respectively. These papers will provide descriptions for the ancillary and external programs, modifications to original targeting strategies required by evaluation of their effectiveness, and modifications of the field plan as required by weather gains or losses. We include all targeting information using flags and also provide input catalogs on the SAS. 

 APOGEE-2 scientific goals are implemented in a three-tier strategy, where individual programs aimed at specific science goals are classified as core, goal, or ancillary. The core programs produce a systematic exploration of the major components of the bulge, disk, and halo and are given the highest priority for implementation. The goal programs have more focused science goals, for example follow-up of Kepler Objects of Interest (KOIs), and are implemented as a secondary priority. Ancillary programs are implemented at the lowest priority; such programs were selected from a competitive proposal process and have only been implemented for APOGEE-2N. Generally, the APOGEE-2N and APOGEE-2S survey science are implemented in the same manner. 

In addition to a target selection analogous to that for the Northern observations, APOGEE-2S includes External Programs selected by the Chilean National Time Allocation Committee (CNTAC) or the Observatories of the Carnegie Institution for Science (OCIS) and led by individual scientists (or teams) who can be external to the SDSS-IV collaboration. External programs can be ``contributed", or proprietary; contributed data are processed through the normal APOGEE data reduction pipelines and are released along with other APOGEE data whereas proprietary programs are not necessarily processed through the standard pipelines or released with the public data releases\footnote{To date all External Programs have been ``contributed" so there are no proprietary external programs.}. The selection of external program targets does not follow the standard APOGEE survey criteria in terms of S/N or even source catalogs; the scientists involved were able to exercise great autonomy in target selection (e.g., no implementation of color cuts). External programs are implemented as classical observing programs with observations only occurring for a given program on nights assigned to it.

The APOGEE portion of DR16 includes 437,485 unique stars. Among the unique stars, 308,000 correspond to core science targets, 112,000 to goal science targets, 13,000 to ancillary APOGEE-2N program targets, and 37,000 to APOGEE-2S external program targets. These numbers add up to more than 437,485 due to some stars being in multiple categories. 

\subsection{APOGEE DR16 Data Products}
The basic procedure for processing and analysis of APOGEE
data is similar to that of DR14 data \citep{2018ApJS..235...42A,Holtzman2018}, but a few notable differences are
highlighted here. Full details, including verification analyses, are presented in \citet{Jonsson2020}.

\subsubsection{Spectral Reduction and Radial Velocity Determinations}

\citet{2015AJ....150..173N} describes the reduction procedure for APOGEE
data.  While the basic reduction steps for DR16 were the same as described there, improvements were
implemented in the handling of bad pixels, flat fielding, and
wavelength calibration, all of which were largely motivated by small differences between the data produced by the APOGEE-S and APOGEE-N instruments. As an improvement over DR14, an attempt was
made to provide rough relative flux calibration for the spectra.
This was achieved by using observations of hot stars on the fiber plug plate
for which the spectral energy distribution are known.

Radial velocities were determined, as in DR14, using cross-correlation against a reference grid,
but a new synthetic grid was calculated for the reference grid,
using the same updated models that were used for the derivation
of stellar parameters and abundances (see \S \ref{details} for details). No constraint was placed on the effective temperature range
of the synthetic grid based on the $J-K$ color; DR14 used such a 
constraint which led to a few issues with bad radial velocities. Therefore DR16 improves on this.

For the faintest stars in DR16, especially those in dwarf
spheroidal galaxies, the individual visit spectra can have
low $S/N$, and, as a result, the radial velocity determination fails. In many,
but not all cases, such objects are flagged as having bad or 
suspect RV combination. Users who are working with data for
stars with $H>14.5$ need to be very careful with these data,
as incorrect RVs leads to incorrect spectral combination, which
invalidates any subsequent analysis. We intend to remedy this problem in the next data release.

\subsubsection{Atmospheric Parameter and Element Abundance Derivations}\label{details}

Stellar parameters and abundances are determined using the
APOGEE Stellar Parameters and Chemical Abundance Pipeline (ASPCAP,
\citealt{garciaperez2016})\footnote{\url{https://github.com/sdss/apogee}}. For DR16, entirely new synthetic
grids were created for this analysis. These grids were based
on a complete set of stellar atmospheres from the MARCS group \citep{Gustafsson2008}
that covers a wide range of \teff, \logg, \feh, \aM,
and [C/M]. Spectral syntheses were performed using the
Turbospectrum code \citep{Plez2012}. The synthesis was done using a revised
APOGEE line-list, which was derived, as before, from matching
very high resolution spectra of the Sun and Arcturus.
The revised line-list differs from that used previously by the
inclusion of lines from FeH, Ce II, and Nd II, some revisions
in the adopted Arcturus abundances, and a proper handling of
the synthesis of a center-of-disk solar spectrum. Details
on the line-list will be presented in V. Smith et al. (in prep). The synthetic
grid for red giants was calculated with seven dimensions, including [N/M] and 
micro-turbulent velocity, as well as the atmospheric parameters previously listed; the
range for [C/M] and [N/M] was expanded over that used for DR14. For the giants, the [C/Fe] grid was expanded to include -1.25 and -1.50 dex and the [N/Fe] dimension to cover from -0.50 to +1.50 dex.
For dwarfs, an additional dimension was included to account for stellar rotation that included 7 steps (these being $v \sin i$ of 1.5, 3.0, 6.0, 12.0, 24.0, 48.0, and 96.0 km s$^{-1}$).   During the stellar parameter and abundance fits, regions in the
spectrum that were not well matched in the solar and Arcturus
spectra were masked. The full details of the spectral grid derivations will be given in a dedicated paper on the APOGEE DR16 pipeline \citep{Jonsson2020}.

The DR16 analysis improves on the measurement of carbon and nitrogen
abundances in dwarf stars over DR14, as DR16 includes separate [C/M]
and [N/M] dimensions for dwarfs.

As for previous data releases, stellar parameters were determined
by searching for the best fit in the synthetic grid. The method
used to normalize the observed and model spectra was improved from
previous releases, and a new minimization option was adopted in
the {\sc FERRE} code \citep{Allende2006}.\footnote{\url{https://github.com/callendeprieto/ferre}}. More details on these changes are given in \citet{Jonsson2020}.  As in previous releases, after the stellar parameters have been determined, these are held fixed while determining the elemental abundances; for these, only windows in the spectra that are sensitive to the element in question are fit, and only a single relevant abundance dimension of the grid is varied. The windows are chosen based on where our synthetic spectra are sensitive to a given element, and at the same time \emph{not} sensitive to another element in the same abundance dimension. In addition to the elements
measured for DR14, an attempt was made to measure the abundance of
cerium using a single line from \citet{Cunha2017}, but these results show
significant scatter and may be of limited utility.

In previous releases, we derived an internal calibration to the abundances to account for biases as a function of  \teff, but for DR16 no such calibration is applied because, with the modification to the abundance pipeline, the trends with effective temperature for most elements have reduced amplitude as compared with previous data processing. The zero-point scale of the abundances was adjusted so that stars in the solar neighborhood (within 0.5 kpc of the Sun, according to {\it Gaia} parallaxes) with near-solar metallicity ($-0.05< $[M/H]$ <0.05$) are adjusted to have a mean [X/M] = 0. The reason for this choice is discussed in detail in \citet{Jonsson2020}.

The procedure is described in significantly more detail, along
with an assessment of the quality of the stellar parameters
and abundances, in \citet{Jonsson2020}.

\subsection{Data Quality}  

\begin{figure*}
\centering
\includegraphics[angle=0,width=15cm]{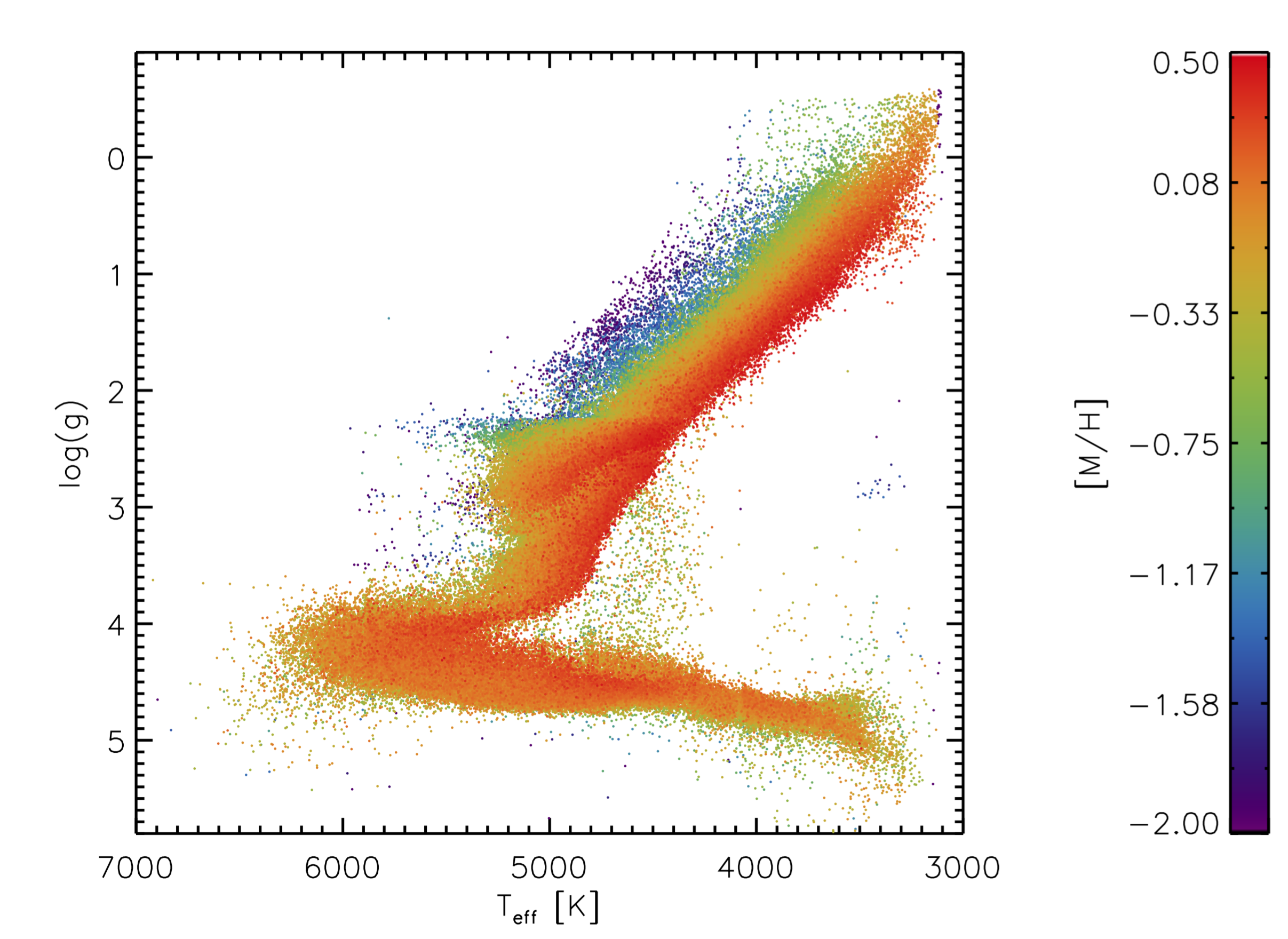}
\caption{Spectroscopic Hertzsprung-Russell diagram, \teff ~versus \logg ~for the main red star sample in APOGEE DR16. The points are color-coded by their total metal content, [M/H]. Dwarf-type stars, those with \logg ~\textgreater 3.7~dex, have calibrated stellar parameters for the first time in DR16. New stellar grids also provide reliable measurements to cooler temperatures than in previous DRs.}
\label{fig:apogeehr}
\end{figure*}

The quality of the DR16 results for radial velocities, stellar parameters, and abundances is 
similar to that of previous APOGEE data releases.  Figure \ref{fig:apogeehr} shows a \teff-\logg ~diagram for the main sample APOGEE 
stars in DR16. The use of MARCS atmosphere models \citep{Gustafsson2008} across the entire \teff-\logg ~range has significantly improved results for cooler giants; previously, Kurucz atmosphere models \citep{Kuruczmodel} were used for the latter stars, and discontinuities were visible at the transition point between MARCS and Kurucz. While the stellar parameters are overall an improvement from previous DRs, we still apply external calibrations to both \logg ~and \teff. These calibrations are discussed fully in \citet{Jonsson2020}, which also describes the features in Figure \ref{fig:apogeehr} in more detail. 

Several fields were observed with both the APOGEE-N and APOGEE-S instruments. Comparing the results, we
find close agreement in the derived stellar parameters and abundances, with mean offsets of 
$\Delta$ \teff $\sim 10$ K, $\Delta$ \logg $\sim 0.02$ dex, and abundance offsets of $<0.02$ dex for
most elements.

 \subsection{APOGEE Value Added Catalogs}
There are six APOGEE-associated VAC's in DR16. A brief description of each VAC and the corresponding publications are given below. They are also listed in Table \ref{table:vac}.

\subsubsection{APOGEE Red Clump Catalog}\label{vac:rc}
DR16 contains the latest version of the APOGEE red-clump (APOGEE-RC) catalog. This catalog is created in the same way as the DR14 version \citep[which is presented in][]{2014ApJ...790..127B}, with the more stringent \logg~cut. The DR16 catalog contains 39,675 unique stars, about 30\% more than in DR14. The red clump stars are cross-matched to {\it Gaia} DR2 \citep{GaiaDR2} by matching (RA, Dec) within a radius of 2 arcsec using the Vizier xmatch service.\footnote{accessed through the {\tt gaia\_tools} code available here: \url{https://github.com/jobovy/gaia\_tools}} We include proper motions through this match.

\subsubsection{APOGEE-\texttt{astroNN}}\label{vac:astroNN}
The APOGEE-\texttt{astroNN} value-added catalog contains the results from applying the \texttt{astroNN} deep-learning code to APOGEE spectra to determine stellar parameters, individual stellar abundances \citep{2019MNRAS.483.3255L}, distances \citep{2019arXiv190208634L}, and ages \citep{2019arXiv190104502M}. Full details of how all of these quantities are determined from the DR16 data are given in \S 2.1 of \citet{2019arXiv190511404B}. In addition, properties of the orbits in the Milky Way (and their uncertainties) for all stars are computed using the fast method of \citet{2018PASP..130k4501M} assuming the \texttt{MWPotential2014} gravitational potential from \citet{2015ApJS..216...29B}. Typical uncertainties in the parameters are 60 K in \teff, 0.2 dex in \logg, 0.05 dex in elemental abundances, 5\,\% in distance, and 30\,\% in age. Orbital properties such as the eccentricity, maximum height above the mid-plane, radial, and vertical action are typically precise to 4--8\,\%.  

\subsubsection{APOGEE-\textit{Joker}}\label{vac:joker}
The APOGEE-\textit{Joker} VAC contains posterior samplings over binary star orbital parameters (i.e., Keplerian orbital elements) for 224,401 stars with three or more APOGEE visit spectra that pass a set of quality cuts as described in \citealt{PriceWhelan2020}).
The samplings are generated using \textit{The Joker}, a custom Monte Carlo sampler designed to handle the very multi-modal likelihood functions that are natural to sparsely-sampled or noisy radial velocity time series \citep{PriceWhelan2017, PriceWhelan2018}.
For some stars, these samplings are unimodal in period, meaning that the data are very constraining and the orbital parameters can be uniquely summarized; in these cases, we provide summary information about the samplings such as the maximum \textit{a posteriori} sample values.

\citet{PriceWhelan2020} describes the resulting catalog from applying of {\it The Joker} to APOGEE DR16. Based on some simple cuts comparing the maximum likelihood posterior sample to the likelihood of a model for each source in which the radial velocities are constant (both quantities are provided in the VAC metadata), we estimate that there are $\gtrsim 25,000$ binary star systems robustly detected by APOGEE (described in \citealt{PriceWhelan2020}, their Section 5). The vast majority of these systems have very poorly constrained orbital parameters, but these posterior samplings are still useful for performing hierarchical modeling of the binary star population parameters (e.g., period distribution and eccentricity parameters) as is demonstrated in \citet{PriceWhelan2020}.

While finalizing the DR16 Value Added Catalog release, we found a bug in the version of {\textit{The Joker} that was used to generate the posterior samplings released in this VAC. This bug primarily impacts long-period orbital parameter samplings, and only for systems with radial velocity measurements that are very noisy or have a short baseline relative to the periods of interest. The samplings for systems with precise data or with many epochs should not be affected. \citet{PriceWhelan2020} describe the this bug in more detail. The VAC will be updated as soon as possible.

\subsubsection{Open Cluster Chemical Abundances and Mapping}\label{vac:occam}

The goal of the Open Cluster Chemical Abundances and Mapping (OCCAM) survey is to create a uniform (same spectrograph, same analysis pipeline) open cluster abundances dataset. We combine proper motion (PM) and radial velocity (RV) measurements from {\it Gaia} DR2 \citep{GaiaDR2} with radial velocity (RV) and metallicity measurements from APOGEE to establish membership probabilities for each star observed by APOGEE in the vicinity of an open cluster. DR16 is the second VAC from the OCCAM survey. We do not impose a minimum number of reliable member stars as in the previous version (released in DR15 \citealt{2019ApJS..240...23A}; and described in detail in \citealt{Donor2018}), but we do enforce a visual quality cut based on each cluster's proper motion (PM) cleaned color-magnitude diagram (CMD). A detailed description of the updated methods is provided in \citet{Donor2020}. The VAC includes 10191 APOGEE stars in the vicinity of 126 open clusters. Average RV, PM, and abundances for reliable ASPCAP elements are provided for each cluster, along with the visual quality determination. Membership probabilities based individually upon RV, PM, and \feh~are provided for each star. The reported cluster PM is from the kernel smoothing routine used to determine cluster membership. Reported RVs and chemical abundances are simply the average value from cluster members; in practice, the uncertainties for chemical abundances are small and show small variation between stars of the same cluster.

\subsubsection{APOGEE DR16 StarHorse Distances and Extinctions}\label{vac:starhorse}

The APOGEE DR16 StarHorse catalog contains updated distance and extinction estimates obtained with the latest version of the StarHorse code \citep{2018MNRAS.476.2556Q, Anders2019}. The DR14 version of these results were published as part of the APOGEE DR14 Distance VAC (\citealt{2018ApJS..235...42A}; Sect. 5.4.3). DR16 results are reported for 388,815 unique stars, based on the following input data: APOGEE DR16 ASPCAP results, broad-band photometry from several sources (PanSTARRS-1, 2MASS, AllWISE), as well as parallaxes from {\it Gaia} DR2 corrected for the zero-point offset of -0.05 mas found by \citet{2019ApJ...878..136Z}. Typical statistical distance uncertainties amount to ~10\% for giant stars and ~3\% for dwarfs, respectively. Extinction uncertainties amount to ~0.07 mag for stars with optical photometry and ~0.17 mag for stars with only infra-red photometry. The APOGEE DR16 StarHorse results are presented in \citet{Quieroz2019}, together with updated results derived using spectroscopic information from other surveys. 

\section{eBOSS: Final Sample Release} 
\label{sec:eboss}

Observations for eBOSS were conducted with the 1000-fiber BOSS spectrograph \citep{Smee2013}
to measure the distance-redshift relation with the
baryon acoustic oscillation (BAO) feature that appears at a scale of roughly 150 Mpc.
The last observations that will contribute to large-scale structure measurements concluded 
on March 1, 2019.  
All eBOSS observations were conducted simultaneously with either TDSS 
observations of variable sources or SPIDERS observations of X-ray sources.

\subsection{eBOSS}

The first generation of SDSS produced a spectroscopic LRG sample \citep{eisenstein01a}
that led to a detection of the BAO feature in the clustering of matter \citep{eisenstein05a}
and the motivation for dedicated large-scale structure surveys within SDSS.
Over the period 2009--2014, BOSS completed a BAO program using more than 1.5 million galaxy
spectra spanning redshifts $0.15<z<0.75$ and more than 150,000 quasars at $z>2.1$
that illuminate the matter density field through the Lyman-$\alpha$ forest.
Operating over the period 2014--2019, eBOSS is the third and final in the series of SDSS large-scale structure surveys.

The eBOSS survey was designed to obtain spectra of four distinct target classes to trace the underlying matter density field
over an interval in cosmic history that was largely unexplored during BOSS.
The LRG sample covers the lowest redshift interval within eBOSS, providing an expansion of the high
redshift tail of the BOSS galaxy sample \citep{reid16a} to a median redshift $z=0.72$.  Galaxy
targets \citep{prakash16a} were selected from imaging catalogs derived from 
Wide-field Infrared Survey Explorer (WISE) \citep[WISE;][]{wright10a}
and SDSS DR13 imaging data.
A new sample of ELG targets covering $0.6<z<1.1$ was observed over the period 2016--2018, leading
to the highest redshift galaxy sample from SDSS.  Galaxy targets were identified using imaging from the Dark Energy Camera
\citep[DECam;][]{flaugher15a}.
The ELG selection \citep{raichoor17a} reaches a median redshift $z=0.85$
and represents the first application of the DECam Legacy Survey data \citep[DECaLS;][]{Dey2019} to
spectroscopic target selection in any large clustering survey.
The quasar sample covers the critical redshift range $0.8 < z < 2.2$ and is
derived from WISE infrared and SDSS optical imaging data \citep{myers15a}.
Finally, new spectra of $z>2.1$ quasars were obtained to enhance the final BOSS
Lyman-$\alpha$ forest measurements \citep{bautista17a,masdesbourboux17a}. A summary of all these target categories, with redshift ranges and numbers, is provided in Table \ref{table:eboss}.

The surface area and target densities of each sample were chosen to maximize sensitivity to the clustering of
matter at the BAO scale.  
The first major clustering result from eBOSS originated from the two-year, DR14 quasar sample.
Using 147,000 quasars, a measurement of the spherically averaged BAO distance at an effective
redshift $z=1.52$ was performed with 4.4\% precision \citep{qso17a}.
The DR14 LRG sample was used successfully to measure the BAO distance scale at
2.6\% precision \citep{bautista17b} while the DR14 high redshift quasar sample led to improved
measurements of BAO in the auto-correlation of the Lyman-$\alpha$ forest \citep{desaintagathe19a}
and the cross-correlation of Lyman-$\alpha$ forest with quasars \citep{blomqvist19a}.
The DR14 samples have also been used to perform measurements of 
redshift-space distortions (RSD) \citep[e.g.][]{zarrouk18a}, tests of inflation \citep[e.g.][]{castorina19a},
and new constraints on the amplitude of matter fluctuations and the
scalar spectral index \citep[e.g.][]{Chabanier19}.

\subsubsection{Scope of eBOSS}

With the completion of eBOSS, the BOSS and eBOSS samples provide six distinct target samples covering
the redshift range $0.2<z<3.5$.  The number of targets for each sample is summarized in Table~\ref{table:eboss}
and the surface density of each sample is shown in Figure~\ref{fig:ebossnz}.

Figure \ref{fig:ebosssky} shows the DR16 eBOSS spectroscopic coverage in Equatorial coordinates.
For comparison, the SDSS-III BOSS coverage is shown in gray.  The programs that define the unique
eBOSS clustering samples are SEQUELS (Sloan Extended Quasar, ELG, and LRG Survey; initiated during SDSS-III; LRG and quasars),
eBOSS LRG+QSO (the primary program in SDSS-IV observing LRGs and Quasi-stellar objects, or QSOs), and ELG (new to DR16). 

\begin{deluxetable}{lcc}
\tablecaption{Main Target Samples in eBOSS and BOSS\label{table:eboss}}
\tablehead{\colhead{Sample} & \colhead{Redshift Range\tablenotemark{a}} &  \colhead{Number}}
\startdata
eBOSS LRGs & $0.6<z<1.0$ & 298762 \\
eBOSS ELGs & $0.6<z<1.1$ & 269889 \\
eBOSS QSOs & $0.8<z<2.2$ & 434820 \\
BOSS ``LOWZ"\tablenotemark{b} & $0.15<z<0.43$ & 343160 \\
BOSS CMASS\tablenotemark{c} & $0.43<z<0.75$ & 862735 \\
BOSS Lyman-$\alpha$ QSOs & $2.2<z<3.5$ & 158917 
\enddata
\tablenotetext{a}{Range used in clustering analysis}
\tablenotetext{b}{The low redshift targets in BOSS}
\tablenotetext{c}{``Constant mass" targets in BOSS}
\end{deluxetable}

\begin{figure}
\centering
\includegraphics[angle=0,width=8.7cm]{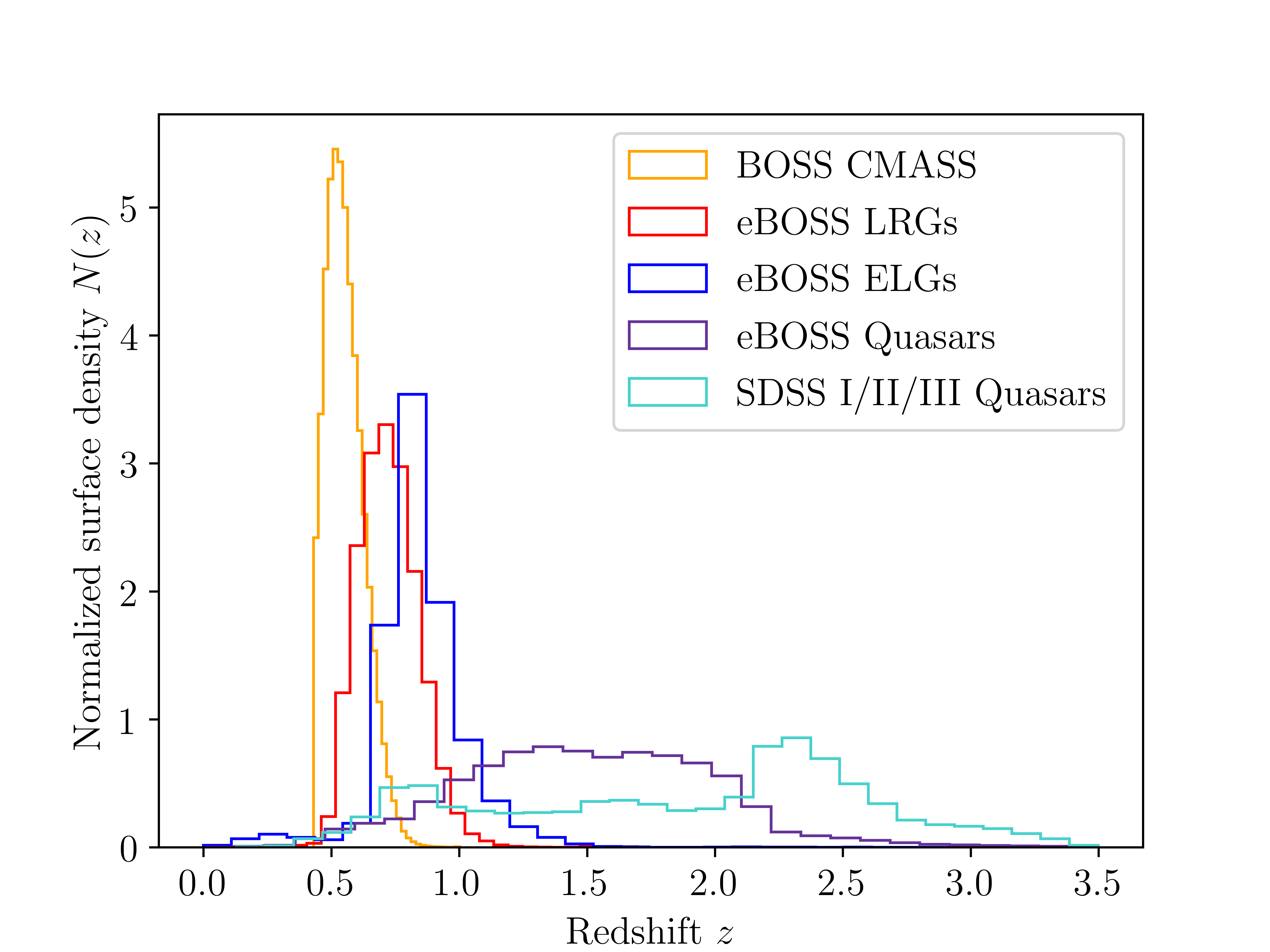}
\caption{The normalized surface density ($N(z)$) of the spectroscopically-confirmed objects used
in the BOSS and eBOSS clustering programs.  The SDSS-I,-II, and -III sample of confirmed
quasars is also presented to demonstrate the gains in the number of quasars that eBOSS
produced over the interval $0.8<z<2.2$.}
\label{fig:ebossnz}
\end{figure}

\begin{figure*}
\centering
\includegraphics[angle=0,width=15cm]{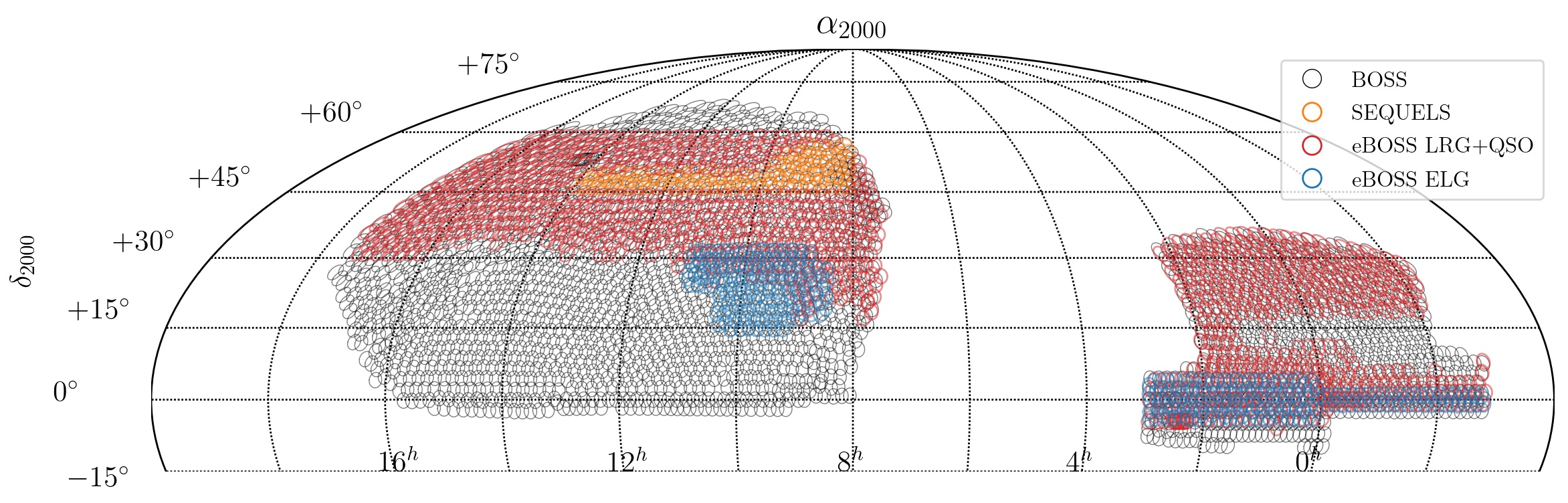}
\caption{DR16 eBOSS spectroscopic coverage in Equatorial coordinates (map centered at RA = 8h.)
Each symbol represents the location of a completed spectroscopic plate scaled to the approximate
field of view. SPIDERS-maximal footprint is the same as BOSS, and SPIDERS-complete is SEQUELS. For more details on SPIDERS coverage see \citep{Comparat2020}.}
\label{fig:ebosssky}
\end{figure*}

\subsubsection{Changes to the eBOSS Spectral Reduction Algorithms}

The data in DR16 were processed with the version {\tt v5\_13\_0} of the
pipeline software \texttt{idlspec2d} \citep{bolton2012,Dawson13a}.
This is the last official version of the software that will be used
for studies of large-scale structure with the SDSS telescope. Table~\ref{tab:pipeline_changes} presents a summary of the major changes in the pipeline during SDSS-IV (eBOSS) and we document the final changes to \texttt{idlspec2d} below.

\begin{deluxetable*}{lll}
\tablecaption{Spectroscopic pipeline major changes \label{tab:pipeline_changes}}
\tablehead{
\colhead{Data Release} & \colhead{\texttt{idlspec2d} version} & \colhead{Major changes} 
}
\startdata
DR12 & \texttt{v5\_7\_0} & Final SDSS-III/BOSS release  \\
DR13 & \texttt{v5\_9\_0} & Adapting software to SDSS-IV/eBOSS data, new unbiased extraction algorithm   \\
DR14 & \texttt{v5\_10\_0} & New unbiased flux correction algorithm, ADR\tablenotemark{a} corrections on individual exposures  \\
DR16 & \texttt{v5\_13\_0} & Improved background fitting in extraction, new stellar templates for flux calibration 
\enddata
\tablenotetext{a}{atmospheric differential refraction} 
\end{deluxetable*}

There were two major changes from DR14 to DR16 to the reduction algorithm.
First, a new set of stellar templates is used for the flux
calibration. This set of templates was produced for the Dark Energy
Spectroscopic Instrument (DESI) pipeline and provided to eBOSS.
These templates reduce residuals in flux calibration relative to previous releases
through improved modeling of spectral lines in the F-stars.
The second major change was in the extraction step, where the background
flux is now fitted prior to the extraction of the flux of individual traces. This modification improved the stability of extraction and removed occasional artifacts
observed in low signal-to-noise spectra. While these changes did not measurably improve the
spectroscopic classification success rates, they represent an improvement
in the overall data quality.

\subsubsection{eBOSS Value Added Catalogs\label{vac:eboss}}
There are two VACs based on eBOSS data which we release in DR16.  These
catalogs offer insight into galaxy physics with eBOSS spectra beyond the core cosmological goals.
The catalogs are described below.

\begin{itemize}
\item {\it Classification eBOSS Emission Line Galaxies:}
 This catalog gives the classification of $0.32<z<0.8$ eBOSS ELGs into four types: star-forming galaxies, composites, Active Galactic Nuclei (AGN) and Low Ionization Nuclear Emission-line Regions (LINERs). It also contains the parameters: [OIII]/H$\beta$, [OII]/H$\beta$, [OIII] line velocity dispersion, stellar velocity dispersion, $u-g$, $g-r$, $r-i$, $i-z$ that are used for classification. The classification is based on a random forest model trained using $z<0.32$ ELGs labeled using standard optical diagnostic diagrams \citep{Zhang2019}. The codes, data and data models are available at \url{https://github.com/zkdtc/MLC\_ELGs} in addition to the standard location for VACs (see \S \ref{sec:access}).
\item {\it FIREFLY Stellar Population Models of SDSS Galaxy Spectra (single fiber):}
We determine the stellar population properties (age, metallicity, dust reddening, stellar mass, and star formation history) for all single fiber spectra classified as galaxies that were published in this release (including those from SDSS-I, II, III and IV). This catalog contains the newly completed samples of eBOSS LRG and eBOSS ELG and will be useful for a variety of studies on galaxy evolution and cosmology \citep[e.g.][]{Bates2019}.
This is an update of the calculation done by \citet{Comparat2017} on the galaxy spectra in DR14 \citep{2018ApJS..235...42A}.
We perform full spectral fitting on individual galaxy spectra using the \textsc{firefly}\footnote{\url{https://github.com/FireflySpectra/firefly_release}} code \citep{Wilkinson_2015,2017MNRAS.465..688G,goddard2017,wilkinson2017} which make use of high spectral resolution stellar population models from \citet{Maraston2011}.
Calculations are carried out using the \citet{Chabrier2003} stellar initial mass function and two input stellar libraries MILES and ELODIE \citep{MILES,MILES_2011,Prugniel2007}.
We publish all catalogs of properties through the SDSS web interfaces (SAS and CAS, see \S \ref{sec:access}) and also make individual best-fit model spectra available through the \textsc{firefly} website\footnote{\url{https://www.sdss.org/dr16/spectro/eboss-firefly-value-added-catalog/}}
\end{itemize}

In the future, we will also present a catalog of more than 800 candidate strong galaxy gravitational lens systems discovered by the presence of higher redshift background emission-lines in eBOSS galaxy spectra (M. Talbot et al. in prep). This Spectroscopic Identification of Lensing Object (SILO) program extends the method of the BOSS Emission-Line Lens Survey~\citep[BELLS;][]{2012ApJ...744...41B} and Sloan Lens ACS~\citep[SLACS;][]{2006ApJ...638..703B} survey to higher redshift, and has recently been applied to the spectroscopic discovery of strongly lensed galaxies in MaNGA~\cite[SILO;][]{2018MNRAS.477..195T}.  The catalog will be released after DR16, but will be based on the DR16 sample.

\subsubsection{Anticipated Cosmology Results from eBOSS}
The final eBOSS BAO and RSD measurements will be presented in a series of independent analyses
for each target class.  The measurements performed with LRG, ELG, and $z<2.2$ quasars
will be performed in configuration space and Fourier space.  Systematic
errors will be assessed through the use of large N-body mock catalogs populated
with galaxies according to a halo occupation distribution prescription
that approximates the observed data, extending the work done in previous data releases \citep[e.g.][]{GIlMarin2018}.  Consensus values of the angular diameter
distance, the Hubble parameter, and $f\sigma_8$
will be provided for each tracer based on the two measurements.
Measurements of the angular diameter distance and the Hubble parameter will be
reported at $z>2.1$ using both the auto-correlation of the final
Lyman-$\alpha$ forest sample and the cross-correlation of the Lyman-$\alpha$ forest
with quasars.  All eBOSS results will be combined with the lower redshift
studies from SDSS and BOSS to offer new constraints on the cosmological model
as was done in the DR11 sample for BOSS \citep{aubourg15a}.

As part of the main cosmological goals of eBOSS, there will be a number of VACs based on the final eBOSS data released in DR16. VACs which are planned and will be publicly released in the future include:
\begin{itemize}
\item {\it Large Scale Structure (from ELGs, LRGs and QSOs).} These large-scale structure (LSS) VACs will be based on all available eBOSS data used for the clustering studies.  Covering the main target classes, this VAC provides the tools to map the three-dimensional structure of the Universe across $0.6 < z<  2.2$ (A. Ross et al. in prep.).
\item {\it Lyman-$\alpha$ Forest Transmission VAC.} This VAC will contain the estimated fluctuations of transmitted flux fraction used for Lyman-$\alpha$ forest BAO measurements.
The catalog will provide the estimates over the Lyman-$\alpha$ and Lyman-$\beta$
rest frame regions of high redshift quasars (H. du Mas des Bourboux in prep.).
\item {\it eBOSS Quasar Catalog.} Beginning with SDSS-I, SDSS has maintained a tradition of releasing a visually-inspected quasar catalog alongside major data releases. The new SDSS-DR16Q catalog (DR16Q; \citealt{lyke2020}) will represent the most recent and largest catalog of known unique quasars within SDSS.
\end{itemize}

\subsection{Reverberation Mapping Program and Other Repeat Spectroscopy}
The SDSS Reverberation Mapping (SDSS-RM; \citealt{2015ApJS..216....4S}) project is a dedicated multi-object reverberation mapping (RM) program that began observations as a part of SDSS-III in January 2014. 
Although not specifically established as a survey within eBOSS, observations of those same targets using
the BOSS spectrograph continued through SDSS-IV.
The SDSS-RM program monitors a sample of 849 quasars in a single $\sim 7\,{\rm deg}^2$ pointing (observed with three plates 7338, 7339 and 7340 with identical targets), with the overall goal of measuring black hole masses via RM in $\sim$100 quasars at a wide range of redshifts (details on the quasar sample itself are provided by \citealt{Shen19a}). During the first season of SDSS-III monitoring, SDSS-RM obtained 32 epochs of SDSS spectroscopy, and has subsequently obtained $\sim 12$ epochs/yr during 2015-2017 and $\sim 6$ epochs/yr during 2018-2020 as part of SDSS-IV. The field has also been monitored photometrically with the Canada-France-Hawaii Telescope (CFHT) and the Steward Observatory Bok telescope in order to increase the observing cadence and the overall yield of RM time-lag measurements. The SDSS-RM field is also coincident with the Pan-STARRS 1 \citep[PS1][]{Kaiser2010} Medium Deep Field MD07, and thus has been monitored photometrically since 2010. Observations with SDSS and the Bok telescope will continue through 2020. 

The program has been largely successful in obtaining RM measurements: \cite{Shen16a} reported several reverberation-mapping measurements from the program after analyzing the first year of spectroscopic data only, and \cite{Li17} measured composite RM signals in the same dataset. \cite{Grier2017} combined the first year of spectroscopy with the first year of photometry and recovered 44 lag measurements in the lowest-redshift subsample using the H$\beta$ emission line. With the additional years of SDSS-IV monitoring included, \cite{Grier19} reported 48 lag measurements using the C{\sc iv} emission line; the addition of another year of SDSS spectroscopy and the inclusion of the PS1 photometric monitoring from 2010--2013 demonstrated the utility of longer time baselines in measuring additional lags (\citealt{Shen19b}). \cite{Homayouni19} measured inter-band continuum lags in many sources, allowing for investigations of accretion-disk properties. Additional studies based off of SDSS-RM data that aim to evaluate and improve RM and black hole-mass measurement methodologies have also been completed (\citealt{Wang19, Li19}). The final SDSS-RM dataset, which will make use of the PS1 monitoring of the SDSS-RM field and seven years of SDSS spectroscopic monitoring, will span more than ten years and allow for the measurement of lags in the highest-luminosity subset of the quasar sample. 

The SDSS-RM dataset is extremely rich and allows for many other types of investigations beyond RM and black-hole masses. The SDSS-RM group has also reported on many other topics, such as studies of quasar host galaxies (\citealt{Shen15b, Matsuoka15, Yue18}), broad absorption-line variability (\citealt{Grier15, Hemler19}), studies of extreme quasar variability (\citealt{Dexter19}) and investigations of quasar emission-line properties (\citealt{Sun15, Denney16a, Shen16b, Denney16b, Sun18}). 
RM observing will continue through 2020 at APO. Building on this program in SDSS-IV an expanded multi-object spectroscopic RM program is included in the Black Hole Mapper program in the upcoming SDSS-V survey post-2020 (see \S \ref{sec:future}).

In addition to the dedicated RM program, there were several fields in SDSS-III and SDSS-IV that were observed multiple times and thus
offer similar potential for time-domain spectroscopic analyses.  Those fields with at least four observations are as follows:
\begin{itemize}
\item {\it Plates 3615 and 3647:} contain the standard BOSS selection of targets.  These two plates have identical science targets
and contain 14 epochs that are classified as ``good'' observations during SDSS-III.
\item {\it Plate 6782:}  contain targets selected to be likely quasars based on variability from multi-epoch imaging data
in Stripe 82 \citep{2000AJ....120.1579Y,Ivezic2007}\footnote{Also see \url{https://classic.sdss.org/dr7/coverage/sndr7.html} for details on Stripe 82 multi-epoch imaging}.  This plate contains four epochs that are classified as ``good'' observations during SDSS-III.
\item {\it Plates 7691 and 10000:} contain a standard eBOSS selection of LRG, quasar, SPIDERS, and TDSS targets.  The two plates
have identical selections and were observed nine times during SDSS-IV.
\item {\it Plate 9414:}  contains ELG targets and TDSS targets from Stripe 82 and was observed four times
to develop higher signal-to-noise spectra that could be used to test the automated redshift classification schemes.  
\end{itemize}
These multi-epoch fields and a few others from BOSS are
described in more detail on the DR16 ``Special Plates'' web page (\url{https://sdss.org/dr16/spectro/special_plates/}).

\subsection{SPIDERS}
\label{sec:spiders}
SPIDERS (Spectroscopic IDentification of EROSITA Sources) is one of two smaller programs conducted within eBOSS. SPIDERS was originally designed as a multi-purpose follow-up program of the Spectrum-Roentgen-Gamma (SRG)/eROSITA all-sky survey \citep{Merloni12,Predehl16}, with the main focus on X-ray selected AGN and clusters of galaxies. Given the delay in the launch of SRG (which took place in July 2019, i.e. after the end of the main eBOSS survey observing) the program was re-purposed to target the X-ray sources from the ROSAT All-Sky Survey \citep[RASS][]{Voges1999,Voges2000} and XMM-Newton \citep[X-ray Multi-mirror Mission][]{Jansen2001AA365L1J}, which will be eventually have their X-ray emission better characterized by eROSITA.

All SPIDERS spectra taken since the beginning of SDSS-IV have targeted either X-ray sources from the revised data reduction of ROSAT \citep[RASS,2RXS][]{Voges1999,Voges2000,Boller16} and XMM-Slew \citep{Saxton08A} catalogs, or red-sequence galaxies in clusters detected by ROSAT (part of the CODEX catalogue, \citealt{Finoguenov2020}) or by XMM \citep[XClass catalogue,][]{Clerc2012}. We define two areas: ``SPIDERS-Maximal'' which correspond to sky area covered by an SDSS legacy or BOSS/eBOSS/SEQUELS plate and ``SPIDERS-Complete'' which corresponds to the area covered by the eBOSS main survey and SEQUELS good plates. SPIDERS-Maximal (Complete) sky area amounts to $10,800$ ($5,350$) $\deg^2$. The sky area corresponding to SPIDERS-Complete is shown in Figure \ref{fig:ebosssky}. 

\subsubsection{SPIDERS Clusters}
In this section we describe the DR16 target selection, data scope, and VACs related to X-ray clusters. 
In DR16, 2,740 X-ray selected clusters (out of a total of 4,114) were spectroscopically confirmed by SPIDERS observing over the SPIDERS-Complete area. 
This constitutes the largest X-ray cluster spectroscopic sample ever build. 
It forms the basis of multiple studies of structure formation on cosmological times \citep{Furnell2018MNRAS4784952F,Erfanianfar2019arXiv190801559E}. 

The majority of SPIDERS clusters targets are galaxies selected via the red-sequence technique around candidate X-ray galaxy clusters \citep{rykoff2012, rykoff2014}. 
These systems were found by filtering X-ray photon over-densities in RASS with an optical cluster finder tool using SDSS photometry. 
The target selection process for these targets is described fully in \citet{Clerc2016}. 
The corresponding target bits and target classes are fully described in the SDSS DR14 data release \citep{2018ApJS..235...42A}. 
We have also considered several additional SPIDERS cluster target classes which we describe below. 

\subsubsection{SPIDERS Target selection update}

New for DR16 is data from ``chunk {\tt eboss20,26,27}". 
In chunk 20, \texttt{SPIDERS\_RASS\_CLUS} targets are obtained by extending the red-sequence search up to five times the cluster virial radius in CODEX clusters detected through their weak-lensing signature \citep{Shan2014}. 
The virial radius used in the target selection is provided in the value-added catalog. 
Moreover, in chunks 26 and 27, we introduce three new target subclasses, taking advantage of deeper optical datasets that enable cluster member measurements at higher redshifts. 

\begin{itemize}
 \item \texttt{SPIDERS\_CODEX\_CLUS\_CFHT}: Following the procedure described in \citet{brimioulle2013}, pointed Canada France Hawaii Telescope (CFHT)/Megacam observations and CFHT-LS fields provide deep $(u)griz$ photometry. A total of 54 (out of 462 targets) spectra were acquired and are labelled with the bit mask \texttt{EBOSS\_TARGET2 = $2^6$};
 \item \texttt{SPIDERS\_CODEX\_CLUS\_PS1}: A sample of 249 high-redshift ($z_{\lambda}>0.5$) CODEX cluster candidates were searched for red-sequence counterparts in PanStarrs PS1 \citep{flewelling2016} using a custom algorithm. A total of 129 (out of 1142 targets) spectra were acquired, and are labelled with the bit mask \texttt{EBOSS\_TARGET2 = $2^7$};
  \item \texttt{SPIDERS\_CODEX\_CLUS\_DECALS}: These targets are output of a custom red-sequence finder code applied to DeCALS photometric data\footnote{http://legacysurvey.org/decamls/} \citep[5th data release][]{Dey2019}. A total of 48 spectra (out of 495 targets) were acquired and are labelled with the bit mask \texttt{EBOSS\_TARGET2 = $2^8$}.
\end{itemize}

\subsubsection{SPIDERS Galaxies and redshifts}
In the SPIDERS-Complete area, a total of 48,013 galaxy redshifts (observed by SDSS-I to IV) are matched to red-sequence galaxy targets, regardless of any actual membership determination (N. Clerc et al. in prep.) 
Of those, 26,527 are SPIDERS targets specifically. 
The additional redshifts were collected from past SDSS-I, II, III and other eBOSS programs. 
The median $i$-band magnitudes of the 26,527 newly acquired targets are $i_{\rm fiber2}=20.0$ and $i_{\rm cModel}=18.5$. 
The spectra are typical of red, passive galaxies at $0.05 \lesssim z \lesssim 0.7$, displaying characteristic absorption features (Ca H+K, G-band, MgI, NaD, etc.) 
Such magnitude and redshift ranges and the purposely narrow spectral diversity make the automated galaxy redshift determination a straightforward task for the eBOSS pipeline, that is well-optimized in this area of the parameter space \citep{bolton2012}. 
In total, 47,492 redshifts are successfully determined with a \texttt{ZWARNING\_NOQSO = 0}. 
The remaining ($\sim 1\%$) showing a non-zero flag are mainly due to due to unplugged fibers or bad columns on the spectrograph CCD or very low signal to noise; their redshift is not measured. 
Full details on the statistical properties of the sample and in particular the success of redshift determination are given in C. Kirkpatrick et al. (in prep.).

\subsubsection{VAC: SPIDERS X-ray clusters catalog for DR16}\label{vac:clusters}
Within the SPIDERS-Complete area, 2,740 X-ray clusters showing a richness $\lambda_{\rm OPT} > 10$ were spectroscopically validated based on galaxy redshift data from SDSS-I to -IV in their red-sequence. 
The richness, $\lambda_{\rm OPT}$, is defined as the sum of the membership probability of every galaxy in the cluster field. 
It was measured by the redmapper algorithm \citep{rykoff2012}. 
A total of 32,326 valid redshifts were associated with these galaxy clusters, leading to a median number of 10 redshifts per cluster red sequence. 
The process of this validation is a combination of automatic and manual evaluations (C. Kirkpatrick et al. in prep). 
An automated algorithm performed a preliminary membership assignment and interloper removal based on standard iterative $\sigma$-clipping methods. 
The results of the algorithm were visually inspected by six experienced galaxy cluster observers (eleven different people since the beginning of the survey), ensuring at least two independent inspectors per system. 
A Web-based interface was specifically developed for this purpose: using as a starting point the result of the automated algorithm, the tool allows each inspector to interactively assess membership based on high-level diagnostics and figures \citep[see Figure 16 in][]{Clerc2016}. 
Validation is in most cases a consequence of finding three or more red-sequence galaxies in a narrow redshift window all within the X-ray estimated viral radius, compatible with them all being galaxy cluster members. 
A robust weighted average of the cluster member redshifts, provides the cluster systemic redshift. 

\subsubsection{X-ray point like sources} \label{vac:agn}
Throughout SDSS-IV, the SPIDERS program has been providing spectroscopic observations of ROSAT/RASS and XMMSL1 sources in the BOSS footprint \citep{Dwelly17}. 
In addition to those targeted by SPIDERS, a large number of ROSAT and XMMSL1 sources received spectra  during the SDSS-I/II \citep[in 2000–2008;][]{2000AJ....120.1579Y} and SDSS- III BOSS \citep[in 2009–2014;][]{Eisenstein2011,Dawson13a} surveys. 
By combining the SDSS-I to IV spectra, the spectroscopic completeness achieved for the ROSAT sample is $10,590/21,945=50$\% in the SPIDERS-Complete area. 
It increases to $53$\% when considering only high-confidence X-ray detections, and to $95$\% when considering only sources with high-confidence X-ray detections and optical counterparts with magnitudes in the nominal eBOSS survey limits ($17\le   i_\texttt{mFiber2} \le 22.5$). 
In the SPIDERS-Maximal area, the spectroscopic completeness of the ROSAT sample is lower $17300/40929=42$\% ($45$\%, $62$\% respectively). 

For ROSAT sources, the major difficulty lies in the identification of secure counterparts of the X-ray sources at optical wavelength, given the large positional uncertainties. 
To solve this problem, the Bayesian cross-matching algorithm \textsc{NWAY} \citep{Salvato18a} was used. The priors for this were based on ALLWISE \citep{Cutri2013} infrared (IR) color-magnitude distributions which, at the depth of the 2RXS and XMMSL2 surveys, can distinguish between X-ray emitting and field sources. 
WISE positions were matched to photometric counterparts in SDSS. 
So that for the DR16 value added catalogues, instead of reporting RASS of XMMSL1 measured X-ray fluxes, we report the updates 2RXS and XMMSL2 fluxes. 
\citet{Comparat2020} presents the SPIDERS spectroscopic survey of X-ray point-like sources, and a detailed description of the DR16 value-added catalogues. We summarize it below.

\subsubsection{VACs: Multi-wavelength Properties of RASS and XMMSL AGNs}\label{vac:rass}

In these two VACs, we present the multiwavelength characterization over the SPIDERS-Complete area of two highly complete samples of X-ray sources:
\begin{enumerate}
    \item The ROSAT All-Sky Survey (RASS) X-ray source catalog \citep[2RXS,][]{Boller16}
    \item The XMM-Newton Slew Survey point source catalog \citep[XMMSL2, Version 2,][]{Saxton08A}.
\end{enumerate}

We provide information about the X-ray properties of the sources as well as of their counterparts at longer wavelengths (optical, IR, radio) identified first in the AllWISE IR catalog via a Bayesian cross-matching algorithm \citep{Salvato18a}. 
We complement this with dedicated visual inspection of all the SDSS spectra, providing accurate redshift estimates (with confidence levels based on the inspection) and source classification, beyond the standard eBOSS pipeline results. 
These two VACs supersede the two analogous ones published in DR14.

\subsubsection{VAC: Spectral Properties and Black Hole Mass Estimates 
for SPIDERS DR16 Type 1 AGN}\label{vac:coffey}

This VAC contains optical spectral properties and black hole mass estimates for the DR16 sample of X-ray selected SPIDERS type 1 (optically unobscured) AGN.
This is the DR16 edition of an earlier SPIDERS VAC covering SPIDERS type 1 AGN up to DR14,
which was presented by \citet{Coffey2019} and \citet{2019ApJS..240...23A}. 
The spectral regions around the MgII and $\rm H\beta$ emission lines were fit using a multicomponent model in order to derive 
optical spectroscopic properties as well as derived quantities such as black hole mass estimates and Eddington ratios.

\subsubsection{Future plans for SPIDERS}
In addition to these programs, completed and fully released in DR16, the performance verification data being taken as part of the eROSITA Final Equatorial Field Depth Survey (eFEDS) is currently planned to be available by November 2019 and should consist of 120 deg$^2$ observed to the final eROSITA all-sky survey depth over an equatorial field overlapping with the GAMA09 \citep{Robotham2011} survey window. 
To address at least part of the original goals of SPIDERS (i.e. eROSITA follow-up) within SDSS-IV, we plan to dedicate a special set of twelve special plates for these targets, to be observed in Spring 2020, and released as part of the final seventeenth data release. An extensive eROSITA follow-up program is also planned for the next generation of the survey, SDSS-V \citep[][and see \S \ref{sec:future}]{2017arXiv171103234K} and 4MOST \citep{Finoguenov2019Msngr.175...39F,Merloni2019Msngr.175...42M}.

\subsection{TDSS}
\label{sec:tdss}
TDSS (The Time Domain Spectroscopic Survey), is the second large subprogram of eBOSS, the goal of which is to provide the first large-scale, systematic survey of spectroscopic follow-up to characterize photometric variables. TDSS makes use of the BOSS spectrographs \citep{Smee2013}, using a small fraction (about 5\%) of the optical fibers on eBOSS plates. TDSS observations thus concluded with the end of the main eBOSS survey data collection 1st March 2019, and the 
full and final TDSS spectroscopic data are included in DR16.

There are three main components of TDSS, each now with data collection 
complete:
\begin{enumerate}
\item The primary TDSS spectroscopic targets are selected from their 
variability within Pan-STARRS1 (PS1 multi-epoch imaging photometry, 
and/or from longer-term photometric variability between PS1 and SDSS 
imaging data, see e.g. \citealt{Morganson2015}). TDSS single epoch 
spectroscopy \citep[SES][]{Ruan2016} of these targets establish the 
nature of  the photometric variable (e.g., variable star vs. variable quasar, and subclass, etc.), and in turn often then suggest the character of the underlying variability (e.g., pulsating RR Lyrae vs. flaring 
late-type star vs. cataclysmic variable, etc.). 
More than 108,000 optical spectra of
these TDSS photometric variables have been observed through DR16 (in both eBOSS and the eBOSS pilot program SEQUELS). Adding in similar variables sources fortuitously already 
having optical spectra within the SDSS archives (from SDSS-I,-II or -III), approximately one-third of 
the TDSS variables can be spectroscopically classified as variable stars, 
and the majority of the remaining two-thirds are variable quasars.
\item A sample of 6,500 TDSS spectroscopic fibers were allotted to obtain
repeat spectra of known star and quasar subclasses of unusual and 
special interest, which were anticipated or suspected to exhibit spectroscopic 
variability in few epoch spectroscopy (FES; see e.g.
\citealt{MacLeod2018}). A recent specific example of this category of sources, are 
TDSS spectra of nearly 250 dwarf Carbon stars that provide strong 
evidence of statistical radial velocity variations indicative of 
subclass binarity \citep{Roulston2019}.
\item The more recently initiated TDSS Repeat Quasar Spectroscopy (RQS) 
 program (see \citealt{MacLeod2018}) obtains multi-epoch spectra for 
16,500 known quasars, sampling across a broad range of properties 
including redshift, luminosity, and quasar subclass type. This has a 
larger sample size, and also a greater homogeneity and less a priori 
bias to specific quasar subclasses compared to the TDSS FES program. All RQS targets have at least one earlier epoch of SDSS spectroscopy already 
available in the SDSS archive. The RQS program is designed especially to investigate
quasar spectral variability on multi-year timescales, and in addition to 
its own potential for new discoveries of phenomena such as changing-look 
quasars or broad absorption line (BAL) variability and others, also 
provides a valuable (and timely) resource for planning of yet larger 
scale multi-epoch quasar repeat spectral observations anticipated for 
the SDSS-V Black Hole Mapper program (see \S \ref{sec:future}).
\end{enumerate}

In total, TDSS has selected or co-selected (in the latter case, often
with eBOSS quasar candidate selections) more than 131,000 spectra in 
SDSS-IV that probe spectroscopy in the time-domain. All of these spectra are now being released in DR16.

\section{MaNGA: Value Added Catalogues Only}
\label{sec:manga}

MaNGA continues to observe galaxies at APO and following the end of eBOSS observing, now uses all dark time at APO. Technical papers are available which overview the project \citep{2015ApJ...798....7B}, target selection \citep{Wake2017}, instrumentation \citep{Drory2015}, observing \citep{Law2015,Yan2016survey} and data reduction and calibration strategies \citep{law16,yan16calibration}.  For DR16 there is no new data release of MaNGA data cubes or analysis products; all remaining data will be released in DR17. However two new or updated MaNGA related VACs are provided which we document here. Previously released VACs, which are still available include those which provide stellar masses, morphologies, and neutral hydrogen (HI) followup (for details of DR15 VACs see \citealt{2019ApJS..240...23A}\footnote{DR15 VACs are found at: \url{https://www.sdss.org/dr15/data_access/value-added-catalogs/}}). 

\subsection{Stellar Masses from Principle Component Analysis}\label{vac:pca}
This VAC provides measurements of resolved and total galaxy stellar-masses, obtained from a low-dimensional fit to the stellar continuum: \citet{Pace2019a} documents the method used to obtain the stellar continuum fit and measurements of resolved stellar mass-to-light ratio, and \citet{Pace2019b}  addresses the aggregation into total galaxy stellar-masses, including aperture-correction and accounting for foreground stars. The measurements rely on MaNGA data reduction pipeline (DRP) version \texttt{v2\_5\_3}, data analysis pipeline (DAP) version \texttt{2.3.0}, and \texttt{PCAY} version \texttt{1.0.0}\footnote{\url{https://www.github.com/zpace/pcay}}. The VAC includes maps of stellar mass-to-light ratio and $i$-band luminosity (in solar units), a table of aperture-corrected total galaxy stellar masses, a library of synthetic model spectra, and the resulting low-dimensional basis set.

The low-dimensional basis set used to fit the stellar continuum is generated by performing principal component analysis (PCA) on a library of 40,000 synthetic star-formation histories (SFHs): the SFHs are delayed-$\tau$ models (${\rm SFR} \sim t ~ e^{-t / \tau}$) modulated by infrequent starbursts, sharp cutoffs, and slow rejuvenations (see \citealt{Pace2019a}, Section 3.1.1). Broad priors dictate the possible range in stellar metallicity, attenuation by foreground dust, and uncertain phases of stellar evolution such as blue stragglers and blue horizontal branch stars (see \citealt{Pace2019a}, Section 3.1.2). The system of six principal component spectra (``eigenspectra") is used as a low-dimensional basis set for fitting the stellar continuum. A distribution of stellar mass-to-light ratio is obtained for each MaNGA spaxel (line of sight in a galaxy) by weighting each model spectrum's known mass-to-light ratio by its likelihood given an observed spectrum. The median of that distribution is adopted as the fiducial stellar mass-to-light ratio of a spaxel, and multiplied by the $i$-band luminosity to get an estimate for the stellar mass.

For DR16, $i$-band stellar mass-to-light ratio and $i$-band luminosity maps (both in Solar units) are released. Stellar mass-to-light ratios have been vetted against synthetic spectra, and found to be reliable at median signal-to-noise ratios between $S/N = 2-20$, across a wide range of dust attenuation conditions (optical depth in the range 0--4), and across the full range of realistic stellar metallicities (-2 dex to +0.2 dex), with respect to Solar (see \citealt{Pace2019a}, Section 4.10). Typical ``random" uncertainties are approximately 0.1 dex (including age-metallicity degeneracies and uncertainties induced by imperfect spectrophotometry), and systematic uncertainties induced by choice of training star formation histories could be as high as 0.3 dex, but are believed to be closer to 0.1--0.15 dex (see \citealt{Pace2019a}, Sections 4.10 \& 5).

In addition to resolved maps of stellar mass-to-light ratio and $i$-band luminosity, the VAC includes a catalog of total stellar masses for MaNGA DR16 galaxies. We provide the total mass inside the integral field unit (IFU; after interpolating over foreground stars and other unreliable measurements with the median of its 8 nearest neighbors: see \citealt{Pace2019b}, Section 4). We also supply two aperture corrections intended to account for mass falling outside the spatial grasp of the IFU: the first adopts the median stellar mass-to-light ratio of the outermost 0.5 effective radii, and the second (recommended) adopts a mass-to-light ratio consistent with the $(g - r)$ color of the NSA flux minus the flux in the IFU (see \citealt{Pace2019b}, Section 4). A comparison of these total masses with those from the NASA-Sloan Atlas (NSA; \citealt{blanton2011}) and MPA-JHU\footnote{Max Planck Institute for Astrophysics and the Johns Hopkins University} catalog \citep{Brinchmann2004} is shown in Figure \ref{fig:mangavac}.

\begin{figure}
\centering
\includegraphics[angle=0,width=8.5cm]{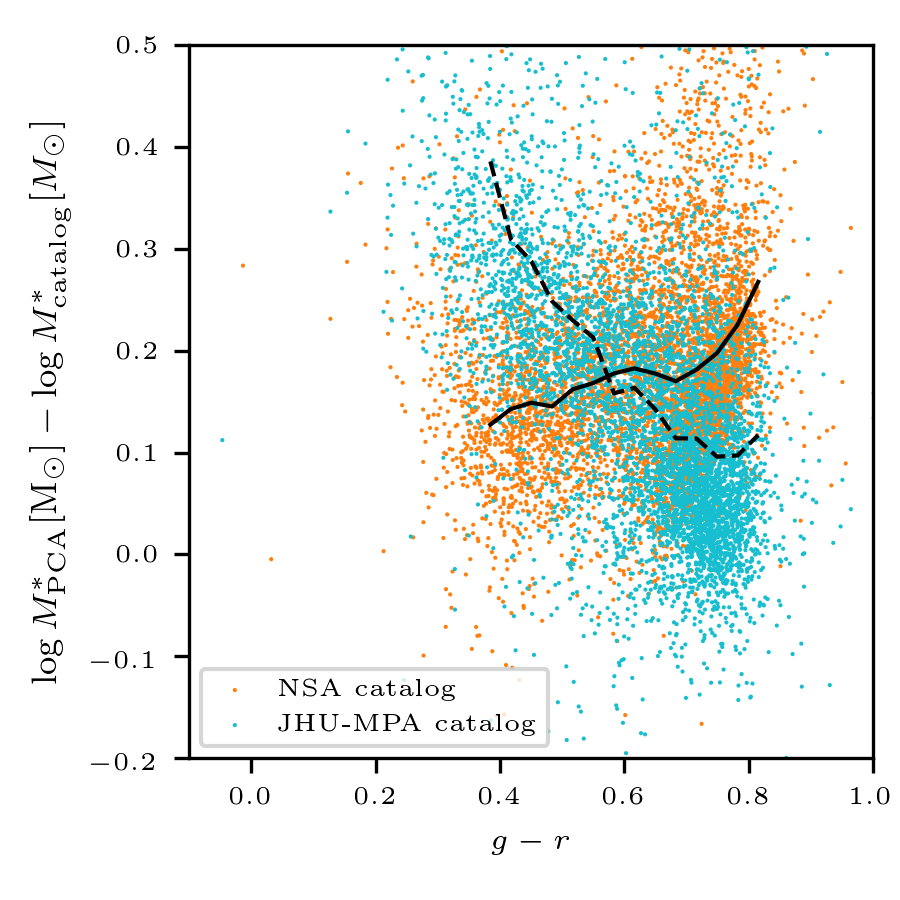}
\caption{A comparison of MaNGA-PCA total stellar masses with NSA (blue points and dashed black line) and MPA-JHU (orange points and solid black line) stellar masses as a function of galaxy $g-r$ colour. The lines show a locally-weighted regression. This plot is reproduced from Figure 6 of \citet{Pace2019b}.}
\label{fig:mangavac}
\end{figure}

\subsection{PawlikMorph Catalog} \label{vac:pawlikmorph}

This catalog provides the shape asymmetry, alongside other standard galaxy morphological related measurements (CAS, Gini, M20, curve of growth radii and S\'ersic fits), based on SDSS DR7 imaging \citep{DR7} using the 8-connected structure detection algorithm described in \citet{Pawlik2016}\footnote{Available from \url{https://github.com/SEDMORPH/PawlikMorph}} to define the edges of the galaxies. We make this available for all galaxies in the MaNGA DR15 release \citep{2019ApJS..240...23A}. The algorithm is specifically designed to identify faint features in the outskirts of galaxies. In this version, stars are not masked prior to creating the 8-connected binary mask, therefore stars lying within the extended light of the galaxies cause incorrect measurements. More than 10\% of objects with unusual measurements have been visually inspected using {\tt Marvin} and SkyServer, and the {\tt WARNINGFLAG} set to 1 for the fraction of these where a star or other problem is identified. Users should not use these measurements, and additionally may wish to visually inspect small samples or outliers to ensure that the sample is appropriate for their science goals.

\section{Conclusions and Future Plans}
\label{sec:future}
This data release, which is the sixteenth over all from SDSS (DR16), is notable for containing the first release of data from Southern hemisphere observing as part of APOGEE-2S and the last release of large scale cosmological redshift-survey data from SDSS (the main program of the eBOSS survey).  DR16 contains no new data from the MaNGA survey. 

SDSS-IV has one final year of operations remaining, and is planning a further one final public data release. That data release, which will be the seventeenth from SDSS overall (DR17), will comprise all remaining data taken by all surveys in SDSS-IV. What follows is a brief summary of the intended contents of DR17: 
\begin{itemize}
\item Due to an accelerated pace of observing in February 2018--1st March 2019, eBOSS has finished observing, and so DR16 is the final data release for both the main eBOSS survey and TDSS. A number of catalogues of redshifts based on eBOSS DR16 spectra have been constructed; these will be released in future. The successful launch of the eROSITA satellite \citep{Predehl_2014_eROSITA} means there will be a small number of addition SPIDERS plates for followup of eROSITA targets, the spectra from which will be released in DR17. 
\item MaNGA has been observing in all remaining dark time from APO since 2nd March 2019, and is on schedule to meet, or slightly exceed its intended goal of 10,000 galaxies. In addition MaNGA has been approved time to observe a subset of ($N\sim$400) galaxies at an exposure time four times deeper than the typical survey. 
\item APOGEE-2 continues to observe from both the Northern (APO) and Southern (LCO) hemisphere. DR16 is the first release of data from the Southern hemisphere, DR17 will be the final release of all APOGEE data from all phases of APOGEE. DR17 will have the complete multi-epoch samples spanning as long as 10 years for some targets, as well as reaching both full depth and coverage in the disk, bulge, and halo programs, and completing large-scale programs to characterize photometric objects of interest in Kepler, K2, and TESS. 
\end{itemize}

\subsection{SDSS-V}

Starting in 2020, after SDSS-IV has ended observations at APO and LCO, the next generation of SDSS will begin --- SDSS-V \citep{2017arXiv171103234K}\footnote{\url{https://www.sdss.org/future}}.  SDSS-V is a multi-epoch spectroscopic survey to observe nearly six million sources using the existing BOSS and APOGEE spectrographs, as well as very large swathes of interstellar medium (ISM) in the Local Group using new optical spectrographs and a suite of small telescopes.  SDSS-V will operate at both APO and LCO, providing the first all-sky ``panoptic'' spectroscopic view of the sky, and will span a wide variety of target types and science goals.

The scientific program is divided into three ``Mappers'': 
\begin{itemize}
\item The {\it Milky Way Mapper} (MWM) is targeting millions of stars with the APOGEE and BOSS spectrographs, ranging from the immediate solar neighborhood to the far side of the Galactic disk and the MW's satellite companions.  The MWM will probe the formation and evolution of the MW, the physics and life-cycles of its stars, and the architecture of multi-star and planetary systems.  
\item The {\it Black Hole Mapper} (BHM) is targeting nearly half a million SMBHs and other X-ray sources (including newly discovered systems from the {\it eROSITA} mission) with the BOSS spectrograph in order to characterize the X-ray sky, measure black hole masses, and trace black hole growth across cosmic time.  
\item Finally, the {\it Local Volume Mapper} (LVM) employs a wide-field optical IFU and new optical spectrographs (with $R \sim 4000$) to map $\sim$2500~deg$^2$ of sky, targeting the ISM and embedded stellar populations in the MW and satellite galaxies.  These maps will reveal the physics of both star formation and the interplay between these stars and the surrounding ISM.
\end{itemize}

SDSS-V builds upon the operational infrastructure and data legacy of earlier SDSS programs, with the inclusion of several key new developments.  Among these are the retirement of the SDSS plug-plate system and the introduction of robotic fiber positioners in the focal planes of both 2.5~m telescopes at APO and LCO.  These focal plane systems (FPS) enable more efficient observing and larger target densities than achievable in previous SDSS surveys.  In addition, the LVM is facilitated by the construction of several $\le$1~meter telescopes at one or both observatories, linked to several new optical spectrographs based on the DESI design \citep{Martini2018}.  SDSS-V continues the SDSS legacy of open data policies and convenient, efficient public data access, with improved data distribution systems to serve its large, diverse, time-domain, multi-object and integral-field data set to the world.
 
After twenty years of Sloan Digital Sky Surveys the data coming out from SDSS-IV in DR16 is making significant contributions to our understanding of the components our Galaxy, galaxy evolution in general and the Universe as a whole. The SDSS-IV project will end with the next data release (DR17), but the future is bright for SDSS with new technology and exciting new surveys coming in SDSS-V.

\section{Acknowledgements}

Funding for the Sloan Digital Sky Survey IV has been provided by
the Alfred P. Sloan Foundation, the U.S. Department of Energy Office of
Science, and the Participating Institutions. SDSS-IV acknowledges
support and resources from the Center for High-Performance Computing at
the University of Utah. The SDSS web site is www.sdss.org.

SDSS-IV is managed by the Astrophysical Research Consortium for the 
Participating Institutions of the SDSS Collaboration including the 
Brazilian Participation Group, the Carnegie Institution for Science, 
Carnegie Mellon University, the Chilean Participation Group, the French Participation Group, Harvard-Smithsonian Center for Astrophysics, 
Instituto de Astrof\'isica de Canarias, The Johns Hopkins University, 
Kavli Institute for the Physics and Mathematics of the Universe (IPMU) / 
University of Tokyo, Korean Participation Group, Lawrence Berkeley National Laboratory, 
Leibniz Institut f\"ur Astrophysik Potsdam (AIP),  
Max-Planck-Institut f\"ur Astronomie (MPIA Heidelberg), 
Max-Planck-Institut f\"ur Astrophysik (MPA Garching), 
Max-Planck-Institut f\"ur Extraterrestrische Physik (MPE), 
National Astronomical Observatories of China, New Mexico State University, 
New York University, University of Notre Dame, 
Observat\'ario Nacional / MCTI, The Ohio State University, 
Pennsylvania State University, Shanghai Astronomical Observatory, 
United Kingdom Participation Group,
Universidad Nacional Aut\'onoma de M\'exico, University of Arizona, 
University of Colorado Boulder, University of Oxford, University of Portsmouth, 
University of Utah, University of Virginia, University of Washington, University of Wisconsin, 
Vanderbilt University, and Yale University.

Co-authorship on SDSS-IV data papers is alphabetical by last name and offered to all collaboration members who have contributed at least 1 month FTE towards any of the surveys during the period up to the end of data collection; and any external collaboration who has contributed at least 1 month FTE to work critical to the data release. 

We would like to thank the Center for Cosmology and AstroParticle Physics (CCAPP) at the Ohio State University for their hospitality during ``DocuBrew" 2019. This event held in August 2019 was the main venue for documentation updates for DR16 (including this paper), was organized by Ashley Ross, Jennifer Johnson and Anne-Marie Weijmans and attended by Rachael Beaton, Joel Brownstein, Brian Cherinka, Kyle Dawson, Sten Hasselquist, Amy Jones, Jade Ho, Karen Masters, Jordan Raddick, Jos\'e S\'anchez-Gallego, Felipe Santana, Michael Talbot (and remotely by Henrik J\"onsson, Julian Bautista, Jon Holtzman, Jennifer Sobeck, Catherine Grier. Johan Comparat, Scott Anderson, Rita Tojeiro, Britt Lundgren and Jesus Pando). Figures \ref{fig:apogeedr16} and \ref{fig:apogeenstars} were made by Christian Hayes. Figure \ref{fig:ebossnz} was made by Ashley Ross, and Figure \ref{fig:ebosssky} by M. Vivek and Julian Baustista. 

This research made use of \textsc{astropy}, a community-developed core \textsc{python} ({\tt http://www.python.org}) package for Astronomy \citep{2013A&A...558A..33A}; \textsc{ipython} \citep{PER-GRA:2007}; \textsc{matplotlib} \citep{Hunter:2007}; \textsc{numpy} \citep{:/content/aip/journal/cise/13/2/10.1109/MCSE.2011.37}; \textsc{scipy} \citep{citescipy}; and \textsc{topcat} \citep{2005ASPC..347...29T}.

\bibliography{references}{}
\bibliographystyle{aasjournal}

\end{document}